\documentclass[aps,prl,reprint,amsmath,amssymb,superscriptaddress,longbibliography]{revtex4-2}
\usepackage{braket}
\usepackage{cancel}

\usepackage[T1]{fontenc}  
\usepackage[utf8]{inputenc}  % Allows input of UTF-8 encoded characters
\usepackage[russian,english]{babel}
\usepackage{graphicx}
\usepackage[normalem]{ulem}
\usepackage{soul}
\usepackage{float} %added by Evyatar
\usepackage[svgnames]{xcolor} %added by Evyatar
\usepackage{comment}
\usepackage[colorlinks, hypertexnames=false, linkcolor=NavyBlue , citecolor= NavyBlue,urlcolor = NavyBlue, breaklinks=true]{hyperref}
\usepackage{verbatim}
\usepackage{esint}
\usepackage{marginnote}
\usepackage{soul}

\newcommand{\RNum}[1]{\uppercase\expandafter{\romannumeral #1\relax}}

\begin{document}

\title{Stability and Relaxation of Electron-Phonon Systems at Strong Coupling}

\author{Noga Bashan}
\affiliation{Department of Condensed Matter Physics, Weizmann Institute of Science, Rehovot 76100, Israel}

\author{Erez Berg}
\affiliation{Department of Condensed Matter Physics, Weizmann Institute of Science, Rehovot 76100, Israel}
\affiliation{Materials Department, University of California Santa Barbara, Santa Barbara 93106 USA}
\affiliation{Department of Electrical and Computer Engineering, University of California, Santa Barbara, CA 93106, USA}

\date{\today}

\begin{abstract}
   %We study the kinetic properties of Migdal-Eliashberg theory (MET) for electron-phonon interactions at strong coupling using the quasiclassical Kadanoff-Baym equation, which is exact in the infinite-bandwidth limit at arbitrary frequencies. Contrary to previous claims of a dynamical instability, we show that the theory remains intrinsically stable at arbitrarily large coupling, and that the previously reported instability is an artifact of an inconsistent use of the gradient approximation.
   
   We study the kinetic properties of Migdal-Eliashberg theory (MET) for electron-phonon interactions at strong coupling. Previously, the system was found to be dynamically unstable at large couplings within a Markovian quantum Boltzmann approximation. Using the quasiclassical Kadanoff-Baym equation, we show that this instability is absent when non-Markovian effects are taken into account.
   We also analyze the long-time relaxation dynamics at strong-couplin, and find a clear imprint of the non-Markovian nature of the dynamics: in an intermediate temperature range between the Debye frequency and the coupling strength, the relaxation rate becomes linear in temperature with Planckian slope, while the leading relaxation mode acquires a large imaginary component, producing strongly oscillatory decay.
   Our results imply that any instability which preempts superconductivity must arise from effects beyond MET, such as phonon softening or polaronic effects at finite electronic bandwidth. 
   This shows that the characteristic phonon frequency is not a fundamental bound on $T_c$, and leaves open a route to higher $T_c$ in multi-orbital electron-phonon systems. 
\end{abstract}
\maketitle

\begin{comment}
    \textbf{Paper structure}
\begin{itemize}
    \item Introduction
    \item KB equations and setup
    \item Results - Response functions and time propagation
    \item Results - relaxation rate
    \item Discussion
\end{itemize}

\textbf{TODO}
\begin{itemize}
    \item Generalize response functions to Ekin-Ekin
    \item negative gammas?
\end{itemize}
\end{comment}

\par\textit{Introduction}---Migdal-Eliashberg theory (MET) of electron-phonon coupling \cite{migdal1958interaction,osti_7354388,eliashberg1961temperature,chubukov2020eliashberg,Sadovskii2022,Marsiglio2020Eliashberg,yuzbashyan2022migdal} is a cornerstone of the modern theory of superconductivity, providing an important step beyond the mean-field BCS description. It enables quantitative calculations of superconducting transition temperatures in various electron-phonon materials \cite{mcmillan1968transition,allen1983theory,carbotte1990properties,kiessling2025bounds2,kiessling2025bounds3}, and provides a natural starting point for generalizations to various unconventional pairing mechanisms \cite{scalapino2012common,abanov2001quantum,abanov2020interplay,abanov2026non,esterlis2026quantum,chowdhury2020unreasonable,inkof2022quantum,tulipman2024solvable,esterlis2021large,kiessling2025bounds1}. Notably, in the strong-coupling regime, MET predicts that the superconducting transition temperature is linearly proportional to the electron-phonon interaction strength \cite{allen1975transition}, suggesting the possibility of parametrically unbounded values of $T_c$.

A central simplification of MET is that many system properties can be formulated in terms of self-consistent equations for electronic correlation functions, while dynamical phonon renormalization effects are parametrically suppressed. Thus, it is applicable non-perturbatively in the interaction strength. However, the strong-coupling regime need not be physically accessible in a generic electron-phonon system. Several effects neglected in this approximation can invalidate the superconducting solution suggested by MET \cite{chubukov2020eliashberg,moussa2006two,esterlis2018bound,CohenAnderson1972}, most notably phonon softening, which can drive a CDW instability, and finite-bandwidth effects, which can favor the formation of polarons or bipolarons \cite{Frohlich01071954,holstein1959studies1,holstein1959studies2,nosarzewski2021superconductivity,chubukov2026breakdown,prokof2026limits,mishchenko2000diagrammatic,prokof1998polaron,esterlis2018breakdown}. These are all effects which are \textit{extrinsic} to MET: the critical interaction threshold and nature of the resulting state  depend on the microscopic details of the model, and are thus nonuniversal phenomena.

Nevertheless, there exist models (e.g. involving multiple electron and phonon flavors) where these effects can be parametrically suppressed \cite{fitzpatrick2014non,Werman2016,werman2017non,hofmann2022heuristic}, extending the range of couplings where MET is valid to the strong-coupling regime. Thus, a natural question is whether MET possesses any generic \textit{intrinsic} instability which may preempt superconductivity? While the answer to this question has been somewhat controversial \cite{yuzbashyan2022breakdown,zhang2023free,yuzbashyan2025instability,semenok2025}, we show that the answer is ``{no}'', and that within the spatially uniform and rotationally symmetric sector considered here, superconductivity is the only intrinsic instability.

To show this, we analyze the kinetics of MET at arbitrary frequencies. This requires generalizing the Quantum Boltzmann Equation (QBE) to the quasiclassical Kadanoff-Baym equation (KBE), which correctly captures the non-Markovian nature of the dynamics. We show that the previously reported instability in the QBE occurs at large imaginary frequency, inconsistent with the gradient approximation on which it relies. To show that the system under consideration is stable, we employ two complementary analyses of its exact description: we diagonalize the linearized KBE's kinetic operator and show that there are no  eigenvalues corresponding to frequencies in the upper half-plane, and we calculate various real-frequency response functions and show that they satisfy stability constraints.

We then use the linearized KBE to study energy-relaxation dynamics at strong coupling. We show that at temperatures above a threshold proportional to the coupling strength the asymptotic relaxation behavior is exactly identical to that predicted by the weak-coupling energy-diffusion model, while below this temperature a different branch dominates the asymptotic behavior. This branch saturates at a ``Planckian'' relaxation rate close to the first Matsubara frequency $\pi T$, and is strongly oscillatory with frequency proportional to interaction strength. This oscillatory late-time decay is a direct dynamical signature of the non-Markovian nature of this regime.

\par\textit{Model and setup}---As a prototypical example of an electron-phonon system, we consider the Holstein model of spinless electrons interacting with Einstein phonons (generalization to spinful electrons is straightforward):
\begin{align}
    H_{\rm el-ph} &= \sum_{k}\xi_k c^\dagger_k c_k + \sum_i \left(\frac{P_i^2}{2M} + \frac{1}{2}M\Omega_0^2 X_i^2 \right) \label{holstein}\\ \nonumber  &+\tilde{g} \sum_i c^\dagger_i c_i X_i 
\end{align}
where $\xi_k=\epsilon_k-\mu$ is the electron's dispersion with $\nu_F$ the DOS at the Fermi level, $M,\Omega_0$ are the phonons' mass and frequency, and $\tilde{g}$ is the interaction strength. We henceforth define a rescaled coupling $g=\sqrt{\nu_F/M}\tilde{g}$ which has units of energy, and for which the dimensionless electron phonon coupling strength is $
    \lambda = {g^2}/{\Omega_0^2}$.
Throughout, we assume that the Fermi-energy and electrons' bandwidth are by far the largest energy scales in the problem, so that 
\begin{equation}
    E_F\sim W\gg T,\Omega_0,g,\lambda\Omega_0,\lambda T
\end{equation}
    
In the finite density Holstein model, the dominant effect of the interaction on the phonons is to soften their frequency, such that $\Omega_0\to \Omega_{\rm eff}=\sqrt{1-\lambda}\Omega_0$, $\lambda\to\lambda_{\rm eff}=\frac{\lambda}{1-\lambda}$ \cite{abrikosov1964methods}. As the bare coupling approaches $\lambda\to1$, the effective frequency vanishes and the effective coupling strength diverges. This motivates interest in the quasielastic limit $\Omega_{\rm eff}/T\to 0$ where the only two relevant energy scales are $g,T$, and MET predicts a superconducting critical temperature $T_c\approx 0.18 \sqrt{\lambda_{\rm eff}}\Omega_{\rm eff}=0.18 g$ \cite{allen1975transition}. Clearly this cannot grow with the coupling indefinitely; when $g=\Omega_0$, $\Omega_{\rm eff}=0$ and the lattice becomes unstable. In practice, in this regime the formation of polarons and bipolarons dramatically suppresses the electronic bandwidth \cite{esterlis2018breakdown,chubukov2026breakdown,prokof2026limits} and MET breaks down.

A convenient route to bypassing such instabilities is to generalize Eq.~\eqref{holstein} to multiple electron and phonon flavors \cite{fitzpatrick2014non,Werman2016,werman2017non}. Consider Eq.~\eqref{holstein} generalized to $N$ fermion flavors $c_{i,a}$ and $N(N+1)/2$ phonon flavors $X_{i,ab}$ (where $a,b=1,\cdots,N$), with modified interaction term
\begin{equation}
    H_{\rm int} = \frac{\tilde{g}}{\sqrt{N}} \sum_i\sum_{a,b=1}^Nc^\dagger_{i,a}c_{i,b} X_{i,ab}.
\end{equation}
This ``matrix-phonon'' structure ensures that both phonon renormalization and vertex corrections are suppressed by a factor of $1/N$ while $\lambda$ remains independent of $N$.  For $N\to\infty$ this guarantees that the phonon frequency and coupling strength are \textit{not} renormalized (i.e. $\Omega_{\rm eff}=\Omega_0,\lambda_{\rm eff}=\lambda$), and justifies the self-consistent one-loop (i.e. no vertex corrections) approximation for the electron self energies, allowing a controlled limit in which the range of validity of MET can be extended to arbitrarily large $\lambda$. 
We mention this large$-N$ construction as motivation for a controlled setting in which non-universal effects can be neglected; the subsequent kinetic analysis applies to the MET equations themselves regardless of the specific Hamiltonian in use.

To study the kinetic properties of this model, we employ the quasiclassical Kadanoff-Baym equation. We consider translationally invariant systems, and project onto the isotropic ($s-$wave) channel, i.e. consider spatially uniform energy dynamics. This is an equation for the Keldysh component of the quasiclassical electron propagator \cite{PrangeKadanoff}, defined as (we mostly follow the conventions of \cite{rammer1986quantum})
\begin{align}
 g^K(t_1,t_2) &\equiv \frac{i}{\pi \nu_F} \int_k G^K(t_1,t_2,\xi_k) \\ \nonumber 
 &\approx \frac{i}{\pi } \int d\xi G^K(t_1,t_2,\xi),  \\ \nonumber G^K(t_1,t_2,\xi_k)&\equiv-i\left<\left[c_k(t_1),c^\dagger_k(t_2)\right]\right>.
\end{align}
Where we have replaced $\int_k\to \nu_F\int d\xi$.
We also define the Moyal product of two-point functions $A,B$ by
\begin{equation}
    A\circ B \space(t_1,t_2) \equiv \int A(t_1,t_3)B(t_3,t_2) dt_3,
\end{equation}
and the Wigner transform of a two-point function $A(t_1,t_2)$ as 
\begin{equation}
    A(t,E) \equiv \int A\left(t+\frac{\tau}{2},t-\frac{\tau}{2}\right)  e^{iE\tau}d\tau. \label{moyal}
\end{equation}
In equilibrium $g^K(t,E)=g^K_{\rm eq}(E)$ is related to the Fermi-Dirac distribution by $g^K_{\rm eq}(E)=2(1-2f_{FD}(E))$, and deviations of $g^K$ encode the nonequilibrium changes in electronic occupations measured by single-body observables. $g^K$'s dependence on the central time $t=\tfrac{t_1+t_2}{2}$ is dictated by the KBE \cite{PrangeKadanoff}:
\begin{align}
    \partial_tg^K = i\left(2\Sigma^K - \Sigma^R\circ g^K + g^K \circ \Sigma^A\right) \label{KBE}
\end{align}
where $\Sigma^{R/A/K}[g^K]$ are the retarded, advanced and Keldysh self energies, all linear functionals of $g^K$ given by MET's  self-consistent one-loop approximation (for further details see the End Matter Eq.~\eqref{self energy}).

For sufficiently slow perturbations, Moyal products can be expanded in gradients in the Wigner basis:
\begin{equation}
    A\circ B = AB +\frac{i}{2}\left(\partial_EA\partial_t B-\partial_t A\partial_E B\right) + \cdots.
\end{equation}
Time-dependent perturbations varying at frequencies much smaller than the system's internal energy scales can be analyzed by truncating the gradient expansion at this order, reducing the KBE to the QBE. For MET near equilibrium all energy derivatives are $O\left(T^{-1}\right)$, and the gradient expansion is controlled for dynamics at frequencies $|\omega|\ll T$.

\par\textit{Stability}---Determination of an instability from a kinetic equation gradient-expanded in time derivatives is subtle, and requires the unstable pole to lie at small enough imaginary frequency. For generic instabilities, where a pole crosses the real frequency axis continuously as a parameter is tuned, this approach is typically consistent. However, it is possible for such a truncation to predict a spurious high-frequency instability which does not exist in the original non-Markovian equation\footnote{An extreme example of this is the non-causal nature of the Abraham-Lorentz force \cite{efimkin2016non}, whose Markovian approximation is non-causal even at arbitrarily weak couplings.}. To see this, consider the simple example of a non-Markovian equation for a single variable $x(t)$ coupled with strength $\alpha$ to a bath whose typical response time is $\tau_{\rm mem}$ :
\begin{equation}
    \dot{x}(t) = -\alpha \int_0^\infty \exp(- s/\tau_{\rm mem}) x(t-s) ds/\tau_{\rm mem}. \label{example}
\end{equation}
The ansatz $x(t)\sim\exp(\gamma t)$ unveils poles at 
$\gamma_\pm = -\left(1 \pm \sqrt{1-4\alpha\tau_{\rm mem}}\right)/2\tau_{\rm mem}$, and since they both have negative real part the system is stable for any value of the coupling $\alpha$. However, if the equation is simplified using the gradient approximation $x(t-s)\approx x(t) - s\dot{x}(t)$, a single pole is found at $\gamma=-\frac{\alpha }{1-\alpha\tau_{\rm mem}}$, which changes sign at $\tau_{\rm mem}\alpha_c=1$ and thus signals an instability. Clearly this contradicts the exact result, and the instability cannot be trusted since $|\gamma(\alpha>\alpha_c)|>1/\tau_{\rm mem}$ is not consistent with the gradient approximation employed. 

As has been pointed out in \cite{yuzbashyan2025instability}, the instability found in the QBE follows similar structure  ($\gamma \propto -\frac{\lambda}{1-\lambda/\lambda_c}$) (and hence the unstable modes are never within the radius of convergence of the gradient expansion). To determine concretely whether the system is stable or not, we extend their analysis to the exact kinetic equation, the KBE. 
%As a preliminary check, we explicitly propagate Eq.~\eqref{KBE} following a quench at $t_1,t_2=0$, using parameters for which the QBE is unstable. As shown in Fig.~\ref{figure 1}(a), the perturbation decays, and the system relaxes to equilibrium smoothly.
We consider the linearization of Eq.~\eqref{KBE} for a small monochromatic perturbation 
\begin{align}
    g^K(t,E)&=g^K_{\rm eq}(E) + \epsilon_\omega(t)\delta g^K(E), \label{perturbation def}\\ \nonumber 
    \epsilon_\omega(t)&=\epsilon_0\exp(-i\omega t) ,
\end{align}
with $|\epsilon_0|\ll1$. The equation can be written in terms of the linearized KBE's kinetic operator $\hat{K}_\omega$
\begin{equation}
    \hat{K}_\omega\delta g^K(E) \equiv\left(-i\omega+\frac{g^2}{T}\hat{C}_\omega\right)\delta g^K(E) = S_\omega(E) \label{lin KBE}
\end{equation}
where $\hat{C}_\omega$ is the dimensionless linear KBE's collision operator at frequency $\omega$ (the right-hand side of Eq.~\eqref{KBE}, linearized with respect to $\epsilon_0$ and divided by $-g^2/T$), and $S_\omega(E)$ stands for a source due to some external drive at frequency $\omega$. Note that within the self consistent one-loop approximation for the self energies (employed in MET), $\hat{C}_\omega$ does not depend explicitly on the interaction strength $g$, only on $\Omega_0,T$. Since we are considering non-Markovian dynamics, the rate of change at a given time is sensitive to the whole history of collisions, and thus $\hat{K}_\omega$ depends nonlinearly on $\omega$. Only for  $|\omega|\ll T$ does it reduce to the linearized QBE kinetic operator \begin{equation}
    \hat{K}^{\rm (QBE)}_\omega \equiv \hat{K}_{\omega\ll T}\approx -i\omega  + \frac{g^2}{T} \left(\hat{C}_{\omega=0}+\omega\partial_\omega \hat{C}_{\omega=0}\right). 
    \label{lin QBE}
\end{equation}

\begin{figure}[t]
\centering
\includegraphics[width=\columnwidth]{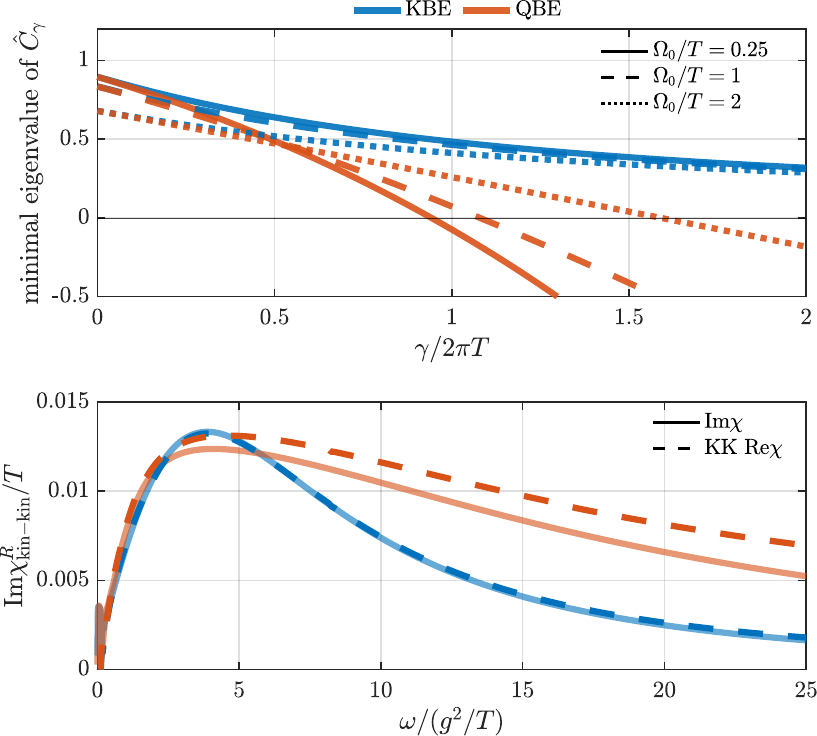} 
\caption{Comparison of stability analyses of the KBE versus QBE. \textbf{Top:} Minimal eigenvalue of the exact KBE (blue) and approximate QBE (orange) collision operators at positive, real, $\gamma$ values, for $\Omega_0/T=0.25,1,2$. As explained in the text, positive values of $\gamma$ for which there exists a negative eigenvalue correspond to unstable behavior at some value of the coupling $g$. This occurs only for the QBE operator. \textbf{Bottom:} Comparison of the imaginary component of $\chi_{\rm kin-kin}^R(\omega)$, calculated using the KBE and QBE, for $\Omega_0/T=1, g^2/T^2=6$ (parameters which lie in the unstable regime of the QBE).
While both agree at frequencies $\omega\lesssim T$, significant deviations appear in the imaginary part at higher frequencies. Dashed lines denote the Kramers-Kronig transformation of the corresponding real parts of $\chi^R$. As is evident from the plot, the KK transformation works well for the exact KBE response and deviates when using the QBE approximation.}
\label{figure 1}
\end{figure}

An instability of MET corresponds to the existence of a zero eigenvalue of $\hat{K}_\omega$ as defined in Eq.~\eqref{lin KBE}  at a frequency in the UHP, i.e. with ${\rm Im}\omega>0$. For convenience we switch to $\gamma = -i\omega$, i.e. consider solutions with time dependence $e^{\gamma t}$.  Here we show that the unstable poles in the QBE response Eq.~\eqref{lin QBE}, which lie at positive, real, values of $\gamma$, disappear in the correct treatment using the KBE (since the unstable modes of the QBE found in \cite{yuzbashyan2025instability} lie in the particle-hole symmetric sector, we restrict to that here. For extension to the antisymmetric sector, as well as to arbitrary complex $\gamma$, see Appendix~\ref{app: stability}).
To identify zeros of $\hat{K}_\gamma$, we search for eigenvalues of $\hat{C}_\gamma$, denoted ${\lambda(\gamma)}$, which satisfy
\begin{equation}
    \gamma = -\frac{g^2}{T}\lambda(\gamma).
    \label{eig condition}
\end{equation}
For positive values of $\gamma$, solutions at \textit{some} value of $g^2/T$ are in one-to-one correspondence with $\gamma$'s for which there exists a negative eigenvalue $\lambda_{\rm min}(\gamma)<0$.

In Fig.~\ref{figure 1}(top) we compare the value of the minimal eigenvalue $\lambda_{\rm min}(\gamma)$, calculated for the KBE and QBE collision operators at multiple values of $\Omega_0/T$. It is apparent that the former does not become negative  anywhere, implying that $\hat{K}_\gamma$ is not singular anywhere in the UHP for any interaction strength. In contrast, the QBE collision operator has eigenvalues which become negative when $|\gamma| \gtrsim \pi T$, lying entirely outside the range of validity of the gradient expansion\footnote{Since $\partial_\gamma \hat{C}_{\gamma=0}$ generally has negative eigenvalues, solutions to Eq.~\eqref{lin QBE} can always be found at large enough values of $|\gamma|$, where it approaches $\hat{K}_\gamma \approx \gamma\left(1 + \frac{g^2}{T}\partial_\gamma \hat{C}_{\gamma=0}\right)$. Thus  the ``critical'' coupling predicted by the QBE is $g_{c}^2/T= 1/\max\left|{\rm eig}\left(\partial_\gamma \hat{C}_{\gamma=0}\right)\right|$. Interestingly, this eigenvalue diverges in the $\Omega_0/T\to 0$ limit.}.

As a complementary test, we study the electron's real-frequency response. For this purpose, we define the low-energy weighted kinetic energy operator,
\begin{equation}
    h_{\Lambda{-\rm kin}} \equiv \sum_k \frac{\xi_k}{1+\left(\xi_k^2/\Lambda^2\right)}c^\dagger_k c_k.
\end{equation}
Here, $\Lambda$ is an artificial regulator intended to curb the linear divergence of response functions at large frequencies (due to Landau damping), which we generically take to be $\Lambda = T$. We calculate the diagonal response function of $h_{\Lambda{-\rm kin}}$, i.e. for a monochromatic drive $H(t)=H_{\rm el-ph}+\epsilon_\omega(t) h_{\Lambda{-\rm kin}}$ we calculate \footnote{This method is equivalent to a calculation of the response function using the Kubo formula, including Bethe-Salpeter re-summation of all ``ladder diagrams''.}
\begin{equation}
    \chi^R_{{\rm kin-kin}}(\omega) \equiv -\left. \frac{\delta\left< h_{\Lambda{-\rm kin}}(t)\right>}{\delta\epsilon_\omega(t)}\right|_{\epsilon_0 = 0}. \label{chi kin kin}
\end{equation}
The exact form of the linearized sources and observables used in this method is presented in the Appendix~\ref{app: linearization}.
In Fig.~\ref{figure 1}(bottom) we compare $\chi^R_{{\rm kin-kin}}(\omega)$ as derived using the KBE Eq.~\eqref{lin KBE} versus QBE Eq.~\eqref{lin QBE}, evaluated for parameters for which the QBE is unstable\footnote{We note that these parameters, corresponding to $\lambda=6,T/g=0.41$ would appear to be in the QBE's stable regime according to Table~1 in the supplementary material of \cite{yuzbashyan2025instability}. We believe this is due to a missing factor of $1/dE^2$ in the discretization of the principal-value term of Eq.~22 of their supplementary material.}. As can be seen, the branch of unstable poles in the QBE leads to broad behavior of the response function at very high frequencies, which is absent in the KBE response. However, there is no clear indication of the instability of the QBE in this parameter range, e.g. a divergence or negative spectral weight. This is due to the fact that the unstable modes do not cross the real frequency axis directly, but rather enter through $|{\rm Im}\omega|=\infty$.

To properly diagnose stability from the real frequency response, we utilize the Kramers-Kronig (KK) relations
\begin{equation}
    \chi^{R}(\omega) = \frac{i}{\pi }{\rm PV} \int \frac{\chi^R(\omega')d\omega'}{\omega-\omega'},
\end{equation}
which hold for any holomorphic function with no poles in the UHP \cite{AltlandSimons2010}. The dashed lines in Fig.~\ref{figure 1}(bottom) are the result of the KK transformation applied to the directly calculated real part of $\chi^R_{\rm kin-kin}(\omega)$. It is evident that when using the QBE, this deviates from the directly calculated imaginary part, exposing the presence of poles in the UHP. On the other hand, for the KBE the two lines agree well, indicative of the stability of the exact solution.

\par
\begin{figure}[t]
\centering
\includegraphics[width=\columnwidth]{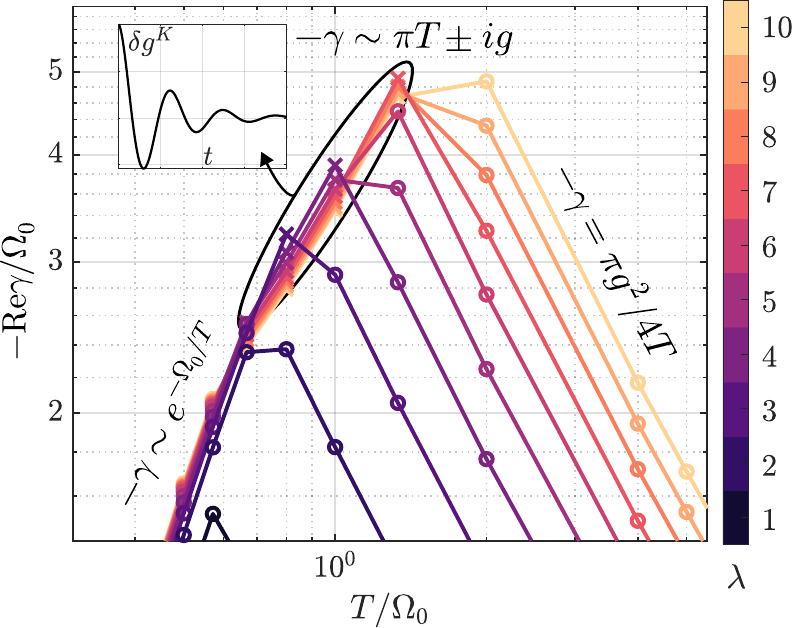} 
\caption{Summary of relaxation regimes at varying values of the dimensionless interaction strength $\lambda$. Points marked with `x' belong to the complex, oscillatory branch.}
\label{figure 3}
\end{figure}

\textit{Relaxation}---Having asserted the stability of the strong-coupling regime of MET, we turn to studying its asymptotic energy-relaxation dynamics\footnote{
Note that we consider relaxation solely in the presence of strong el-ph coupling. A common yet different analysis is of relaxation in the presence of el-el interactions which are significantly stronger than the el-ph interactions, such that the bottleneck for relaxation is the perturbative transfer of energy between electrons and phonons \cite{allen1987theory,SciPostPhys.13.4.095,guo2026phonon}. In that case, at timescales longer than the el-el relaxation time, perturbations to the electrons can be projected to a family of thermal distributions with time dependence carried by the electron's temperature. This agrees with our results at sufficiently small el-ph coupling, as shown in  Appendix~\ref{app: relaxation}. We note that it is not clear a-priori that a similar separation of timescales exists in the case where both el-el and el-ep interactions are sufficiently strong; as we show in the rest of this work, stronger interactions do not necessarily lead to faster relaxation, and the relaxation rate may instead saturate at a ``Planckian'' limit.}. Since we parametrize time dependence as $e^{\gamma t}$, we now search for $\gamma$ values with least negative (nonzero) real part at which $\hat{K}_\gamma$ has a zero eigenvalue (focusing on the particle-hole symmetric sector). 
We plot the result of this analysis, i.e. the slowest $\gamma(T)$, in Fig.~\ref{figure 3}, with different curves corresponding to different values of the dimensionless electron-phonon coupling $\lambda$. 
A thorough derivation of these results, along with a more intuitive characterization of the real solutions, is provided in Appendix~\ref{app: relaxation}.

The behaviour at sufficiently high and low temperatures is as expected: when $T\gg \Omega_0,g$ the system is well described by the energy-diffusion model, which gives $-\gamma=\frac{\pi g^2}{4T}$. This result can be derived exactly at weak couplings using the Boltzmann equation, and appears to hold non-perturbatively in $\lambda$. This is due to the fact that the corresponding eigenmodes can be mapped to plane-waves on the energy axis, with a dispersion that is robust to sufficiently local deformations of $\hat{K}_\gamma$ (see Appendix~\ref{app: relaxation} for more details). 
At low temperatures ($T\lesssim\Omega_0$) $\gamma$ becomes exponentially suppressed by the phonon's thermal factor, and decreases rapidly as $T\to 0$. 

A surprising new result is the emergence of a new branch of eigenvalues which dominates asymptotic relaxation at intermediate temperatures for sufficiently large $\lambda$. This branch is complex, with $-{\rm Re}\gamma \approx \pi T$ and ${\rm Im}\gamma \propto \pm g$, thus it corresponds to a strongly oscillatory relaxation mode, with frequency proportional to the interaction strength and a decaying envelope which only depends on temperature. The range of temperatures where it is the slowest relaxing mode is determined by competition with the high and low temperature branches: it is cut off at low temperatures when $T\approx 0.8 \Omega_0$, and at high temperatures when $T\approx g/2$. Thus, this branch is visible for some intermediate interval of temperatures when $\lambda \gtrsim 2.5$.

\par\textit{Discussion}---Our central result is the determination that the Migdal-Eliashberg saddle can remain locally stable up to arbitrary interaction strengths, at least as far as zero wavevector and angular momentum perturbations are concerned. 
This shows that as a matter of principle, the mediating boson's frequency cannot act as a general bound to the superconducting $T_c$ \cite{hofmann2022heuristic}. Models, such as the large$-N$ construction discussed in the introduction, can host superconductors with $T_c$ up to a significant fraction of the Fermi energy.

    There remains the practical question of the relevance of the large $\lambda$ regime to electron-phonon systems realized in nature. As has been pointed out \cite{esterlis2018bound,semenok2025}, conventional electron-phonon superconductors do not exhibit exceptionally large values of $\lambda$; the record holding experimentally realized materials are amorphous $\rm PbBi$ with $\lambda\approx 3$\cite{chen1971electron} and the hydride $\rm(La,Y)H_{10}$ with $\lambda \approx 3.8$\cite{semenok2021superconductivity}. 
    However, the mechanism stabilizing the large-$N$ construction (spreading $\lambda$ between multiple, frustrated phonon modes) could serve as a useful design principle in the search for new strong coupling electron-phonon superconductors.

Through analyzing the system's relaxation behavior, we have identified a temperature regime existing at sufficiently large couplings (starting around $\lambda\gtrsim2.5$) where relaxation is genuinely non-Markovian, exhibiting oscillations in the slowest relaxing mode. There exist experimental methods which could measure  such dynamics directly \cite{kumar2025absencetwophononquasielasticscattering,barbalas2025energy},  and the aforementioned large-$\lambda$ amorphous compounds could be feasible candidates for observing this behaviour.

On the theoretical level, it is interesting to note that relaxation in this model satisfies a ``Planckian'' bound on equilibration $-{\rm Re}\gamma \lesssim \pi T$ \cite{hartnoll2022colloquium,sachdev1999universal,zaanen2004temperature}. The mechanism behind this saturation is especially transparent within the non-Markovian framework we utilize: for sufficiently large $\lambda$, the system attempts to decay to equilibrium at timescales shorter than its own memory time (which is the thermal timescale $\tau_{\rm mem}\sim (\pi T)^{-1}$). Because it cannot respond to itself quickly enough, it ``over-shoots'' equilibrium, and the increased $\lambda$ leads to oscillations instead of faster relaxation. This behaviour is very generic in non-Markovian dynamics, and is even realized in the toy-model Eq.~\eqref{example}, which passes an exceptional point at $\alpha\tau_{\rm mem}=1/4$ and saturates $-{\rm Re}\gamma_- \leq 1/(2\tau_{\rm mem})$.

It is interesting to investigate how this relates to other similar proposed bounds \cite{qi2026planckian},  whether this generalizes to other models which can be described within a similar framework (e.g. for bosons with a finite lifetime or critical bosons), and how this relates to models with more intricate momentum structure \cite{GuoFluctation}.

\begin{acknowledgments}
The authors are grateful to B. Altshuler, A. V. Chubukov, P. Emanuel, S. A. Kivelson, D. Maslov, E. Tulipman and E. Yuzbashyan for helpful discussions. This work was supported by the Simons Foundation Collaboration on New Frontiers in Superconductivity (Grant SFI-MPS-NFS-00006741-03), an NSF-BSF Award No. DMR-2310312, and CRC 183 of the Deutsche Forschungsgemeinschaft (Project C02). This work was performed in part at the Aspen Center for Physics, which is supported by National Science Foundation grant PHY-2210452.

\end{acknowledgments}

\bibliography{Eth_updated_formatting}

\begin{thebibliography}{69}%
\makeatletter
\providecommand \@ifxundefined [1]{%
 \@ifx{#1\undefined}
}%
\providecommand \@ifnum [1]{%
 \ifnum #1\expandafter \@firstoftwo
 \else \expandafter \@secondoftwo
 \fi
}%
\providecommand \@ifx [1]{%
 \ifx #1\expandafter \@firstoftwo
 \else \expandafter \@secondoftwo
 \fi
}%
\providecommand \natexlab [1]{#1}%
\providecommand \enquote  [1]{``#1''}%
\providecommand \bibnamefont  [1]{#1}%
\providecommand \bibfnamefont [1]{#1}%
\providecommand \citenamefont [1]{#1}%
\providecommand \href@noop [0]{\@secondoftwo}%
\providecommand \href [0]{\begingroup \@sanitize@url \@href}%
\providecommand \@href[1]{\@@startlink{#1}\@@href}%
\providecommand \@@href[1]{\endgroup#1\@@endlink}%
\providecommand \@sanitize@url [0]{\catcode `\\12\catcode `\$12\catcode `\&12\catcode `\#12\catcode `\^12\catcode `\_12\catcode `\%12\relax}%
\providecommand \@@startlink[1]{}%
\providecommand \@@endlink[0]{}%
\providecommand \url  [0]{\begingroup\@sanitize@url \@url }%
\providecommand \@url [1]{\endgroup\@href {#1}{\urlprefix }}%
\providecommand \urlprefix  [0]{URL }%
\providecommand \Eprint [0]{\href }%
\providecommand \doibase [0]{https://doi.org/}%
\providecommand \selectlanguage [0]{\@gobble}%
\providecommand \bibinfo  [0]{\@secondoftwo}%
\providecommand \bibfield  [0]{\@secondoftwo}%
\providecommand \translation [1]{[#1]}%
\providecommand \BibitemOpen [0]{}%
\providecommand \bibitemStop [0]{}%
\providecommand \bibitemNoStop [0]{.\EOS\space}%
\providecommand \EOS [0]{\spacefactor3000\relax}%
\providecommand \BibitemShut  [1]{\csname bibitem#1\endcsname}%
\let\auto@bib@innerbib\@empty
%</preamble>
\bibitem [{\citenamefont {Migdal}(1958)}]{migdal1958interaction}%
  \BibitemOpen
  \bibfield  {author} {\bibinfo {author} {\bibfnamefont {A.~B.}\ \bibnamefont {Migdal}},\ }\bibfield  {title} {\bibinfo {title} {Interaction between electrons and lattice vibrations in a normal metal},\ }\href@noop {} {\bibfield  {journal} {\bibinfo  {journal} {Sov. Phys. JETP}\ }\textbf {\bibinfo {volume} {7}},\ \bibinfo {pages} {996} (\bibinfo {year} {1958})}\BibitemShut {NoStop}%
\bibitem [{\citenamefont {Eliashberg}(1960)}]{osti_7354388}%
  \BibitemOpen
  \bibfield  {author} {\bibinfo {author} {\bibfnamefont {G.~M.}\ \bibnamefont {Eliashberg}},\ }\bibfield  {title} {\bibinfo {title} {Interactions between electrons and lattice vibrations in a superconductor},\ }\href {https://www.osti.gov/biblio/7354388} {\bibfield  {journal} {\bibinfo  {journal} {Sov. Phys. JETP}\ }\textbf {\bibinfo {volume} {11}},\ \bibinfo {pages} {696} (\bibinfo {year} {1960})}\BibitemShut {NoStop}%
\bibitem [{\citenamefont {Eliashberg}(1961)}]{eliashberg1961temperature}%
  \BibitemOpen
  \bibfield  {author} {\bibinfo {author} {\bibfnamefont {G.~M.}\ \bibnamefont {Eliashberg}},\ }\bibfield  {title} {\bibinfo {title} {Temperature {Green}'s function for electrons in a superconductor},\ }\href@noop {} {\bibfield  {journal} {\bibinfo  {journal} {Sov. Phys. JETP}\ }\textbf {\bibinfo {volume} {12}},\ \bibinfo {pages} {1000} (\bibinfo {year} {1961})}\BibitemShut {NoStop}%
\bibitem [{\citenamefont {Chubukov}\ \emph {et~al.}(2020)\citenamefont {Chubukov}, \citenamefont {Abanov}, \citenamefont {Esterlis},\ and\ \citenamefont {Kivelson}}]{chubukov2020eliashberg}%
  \BibitemOpen
  \bibfield  {author} {\bibinfo {author} {\bibfnamefont {A.~V.}\ \bibnamefont {Chubukov}}, \bibinfo {author} {\bibfnamefont {A.}~\bibnamefont {Abanov}}, \bibinfo {author} {\bibfnamefont {I.}~\bibnamefont {Esterlis}},\ and\ \bibinfo {author} {\bibfnamefont {S.~A.}\ \bibnamefont {Kivelson}},\ }\bibfield  {title} {\bibinfo {title} {{Eliashberg} theory of phonon-mediated superconductivity—when it is valid and how it breaks down},\ }\href@noop {} {\bibfield  {journal} {\bibinfo  {journal} {Annals of Physics}\ }\textbf {\bibinfo {volume} {417}},\ \bibinfo {pages} {168190} (\bibinfo {year} {2020})}\BibitemShut {NoStop}%
\bibitem [{\citenamefont {Sadovskii}(2022)}]{Sadovskii2022}%
  \BibitemOpen
  \bibfield  {author} {\bibinfo {author} {\bibfnamefont {M.~V.}\ \bibnamefont {Sadovskii}},\ }\bibfield  {title} {\bibinfo {title} {Limits of {Eliashberg} theory and bounds for superconducting transition temperature},\ }\href {https://doi.org/10.3367/UFNe.2021.05.039007} {\bibfield  {journal} {\bibinfo  {journal} {Physics-Uspekhi}\ }\textbf {\bibinfo {volume} {65}},\ \bibinfo {pages} {724} (\bibinfo {year} {2022})}\BibitemShut {NoStop}%
\bibitem [{\citenamefont {Marsiglio}(2020)}]{Marsiglio2020Eliashberg}%
  \BibitemOpen
  \bibfield  {author} {\bibinfo {author} {\bibfnamefont {F.}~\bibnamefont {Marsiglio}},\ }\bibfield  {title} {\bibinfo {title} {{Eliashberg} theory: A short review},\ }\href {https://doi.org/10.1016/j.aop.2020.168102} {\bibfield  {journal} {\bibinfo  {journal} {Annals of Physics}\ }\textbf {\bibinfo {volume} {417}},\ \bibinfo {pages} {168102} (\bibinfo {year} {2020})}\BibitemShut {NoStop}%
\bibitem [{\citenamefont {Yuzbashyan}\ and\ \citenamefont {Altshuler}(2022{\natexlab{a}})}]{yuzbashyan2022migdal}%
  \BibitemOpen
  \bibfield  {author} {\bibinfo {author} {\bibfnamefont {E.~A.}\ \bibnamefont {Yuzbashyan}}\ and\ \bibinfo {author} {\bibfnamefont {B.~L.}\ \bibnamefont {Altshuler}},\ }\bibfield  {title} {\bibinfo {title} {{Migdal-Eliashberg} theory as a classical spin chain},\ }\href@noop {} {\bibfield  {journal} {\bibinfo  {journal} {Physical Review B}\ }\textbf {\bibinfo {volume} {106}},\ \bibinfo {pages} {014512} (\bibinfo {year} {2022}{\natexlab{a}})}\BibitemShut {NoStop}%
\bibitem [{\citenamefont {McMillan}(1968)}]{mcmillan1968transition}%
  \BibitemOpen
  \bibfield  {author} {\bibinfo {author} {\bibfnamefont {W.~L.}\ \bibnamefont {McMillan}},\ }\bibfield  {title} {\bibinfo {title} {Transition temperature of strong-coupled superconductors},\ }\href@noop {} {\bibfield  {journal} {\bibinfo  {journal} {Physical Review}\ }\textbf {\bibinfo {volume} {167}},\ \bibinfo {pages} {331} (\bibinfo {year} {1968})}\BibitemShut {NoStop}%
\bibitem [{\citenamefont {Allen}\ and\ \citenamefont {Mitrovi{\'c}}(1983)}]{allen1983theory}%
  \BibitemOpen
  \bibfield  {author} {\bibinfo {author} {\bibfnamefont {P.~B.}\ \bibnamefont {Allen}}\ and\ \bibinfo {author} {\bibfnamefont {B.}~\bibnamefont {Mitrovi{\'c}}},\ }\bibfield  {title} {\bibinfo {title} {Theory of superconducting {$T_c$}},\ }\href@noop {} {\bibfield  {journal} {\bibinfo  {journal} {Solid State Physics}\ }\textbf {\bibinfo {volume} {37}},\ \bibinfo {pages} {1} (\bibinfo {year} {1983})}\BibitemShut {NoStop}%
\bibitem [{\citenamefont {Carbotte}(1990)}]{carbotte1990properties}%
  \BibitemOpen
  \bibfield  {author} {\bibinfo {author} {\bibfnamefont {J.~P.}\ \bibnamefont {Carbotte}},\ }\bibfield  {title} {\bibinfo {title} {Properties of boson-exchange superconductors},\ }\href@noop {} {\bibfield  {journal} {\bibinfo  {journal} {Reviews of Modern Physics}\ }\textbf {\bibinfo {volume} {62}},\ \bibinfo {pages} {1027} (\bibinfo {year} {1990})}\BibitemShut {NoStop}%
\bibitem [{\citenamefont {Kiessling}\ \emph {et~al.}(2025{\natexlab{a}})\citenamefont {Kiessling}, \citenamefont {Altshuler},\ and\ \citenamefont {Yuzbashyan}}]{kiessling2025bounds2}%
  \BibitemOpen
  \bibfield  {author} {\bibinfo {author} {\bibfnamefont {M.~K.-H.}\ \bibnamefont {Kiessling}}, \bibinfo {author} {\bibfnamefont {B.~L.}\ \bibnamefont {Altshuler}},\ and\ \bibinfo {author} {\bibfnamefont {E.~A.}\ \bibnamefont {Yuzbashyan}},\ }\bibfield  {title} {\bibinfo {title} {Bounds on {$T_c$} in the {Eliashberg} theory of superconductivity. {II}: Dispersive phonons},\ }\href@noop {} {\bibfield  {journal} {\bibinfo  {journal} {Journal of Statistical Physics}\ }\textbf {\bibinfo {volume} {192}},\ \bibinfo {pages} {94} (\bibinfo {year} {2025}{\natexlab{a}})}\BibitemShut {NoStop}%
\bibitem [{\citenamefont {Kiessling}\ \emph {et~al.}(2025{\natexlab{b}})\citenamefont {Kiessling}, \citenamefont {Altshuler},\ and\ \citenamefont {Yuzbashyan}}]{kiessling2025bounds3}%
  \BibitemOpen
  \bibfield  {author} {\bibinfo {author} {\bibfnamefont {M.~K.-H.}\ \bibnamefont {Kiessling}}, \bibinfo {author} {\bibfnamefont {B.~L.}\ \bibnamefont {Altshuler}},\ and\ \bibinfo {author} {\bibfnamefont {E.~A.}\ \bibnamefont {Yuzbashyan}},\ }\bibfield  {title} {\bibinfo {title} {Bounds on {$T_c$} in the {Eliashberg} theory of superconductivity. {III}: Einstein phonons},\ }\href@noop {} {\bibfield  {journal} {\bibinfo  {journal} {Journal of Statistical Physics}\ }\textbf {\bibinfo {volume} {192}},\ \bibinfo {pages} {93} (\bibinfo {year} {2025}{\natexlab{b}})}\BibitemShut {NoStop}%
\bibitem [{\citenamefont {Scalapino}(2012)}]{scalapino2012common}%
  \BibitemOpen
  \bibfield  {author} {\bibinfo {author} {\bibfnamefont {D.~J.}\ \bibnamefont {Scalapino}},\ }\bibfield  {title} {\bibinfo {title} {A common thread: The pairing interaction for unconventional superconductors},\ }\href@noop {} {\bibfield  {journal} {\bibinfo  {journal} {Reviews of Modern Physics}\ }\textbf {\bibinfo {volume} {84}},\ \bibinfo {pages} {1383} (\bibinfo {year} {2012})}\BibitemShut {NoStop}%
\bibitem [{\citenamefont {Abanov}\ \emph {et~al.}(2001)\citenamefont {Abanov}, \citenamefont {Chubukov},\ and\ \citenamefont {Schmalian}}]{abanov2001quantum}%
  \BibitemOpen
  \bibfield  {author} {\bibinfo {author} {\bibfnamefont {A.}~\bibnamefont {Abanov}}, \bibinfo {author} {\bibfnamefont {A.~V.}\ \bibnamefont {Chubukov}},\ and\ \bibinfo {author} {\bibfnamefont {J.}~\bibnamefont {Schmalian}},\ }\bibfield  {title} {\bibinfo {title} {Quantum-critical superconductivity in underdoped cuprates},\ }\href@noop {} {\bibfield  {journal} {\bibinfo  {journal} {EPL (Europhysics Letters)}\ }\textbf {\bibinfo {volume} {55}},\ \bibinfo {pages} {369} (\bibinfo {year} {2001})}\BibitemShut {NoStop}%
\bibitem [{\citenamefont {Abanov}\ and\ \citenamefont {Chubukov}(2020)}]{abanov2020interplay}%
  \BibitemOpen
  \bibfield  {author} {\bibinfo {author} {\bibfnamefont {A.}~\bibnamefont {Abanov}}\ and\ \bibinfo {author} {\bibfnamefont {A.~V.}\ \bibnamefont {Chubukov}},\ }\bibfield  {title} {\bibinfo {title} {Interplay between superconductivity and non-{Fermi} liquid at a quantum critical point in a metal. {I}. the $\gamma$ model and its phase diagram at {$T=0$}: The case {$0<\gamma<1$}},\ }\href@noop {} {\bibfield  {journal} {\bibinfo  {journal} {Physical Review B}\ }\textbf {\bibinfo {volume} {102}},\ \bibinfo {pages} {024524} (\bibinfo {year} {2020})}\BibitemShut {NoStop}%
\bibitem [{\citenamefont {Abanov}\ and\ \citenamefont {Chubukov}(2026)}]{abanov2026non}%
  \BibitemOpen
  \bibfield  {author} {\bibinfo {author} {\bibfnamefont {A.~G.}\ \bibnamefont {Abanov}}\ and\ \bibinfo {author} {\bibfnamefont {A.~V.}\ \bibnamefont {Chubukov}},\ }\bibfield  {title} {\bibinfo {title} {Non-{BCS} pairing by a singular dynamical interaction},\ }\href@noop {} {\bibfield  {journal} {\bibinfo  {journal} {arXiv preprint arXiv:2606.21731}\ } (\bibinfo {year} {2026})}\BibitemShut {NoStop}%
\bibitem [{\citenamefont {Esterlis}\ and\ \citenamefont {Schmalian}(2026)}]{esterlis2026quantum}%
  \BibitemOpen
  \bibfield  {author} {\bibinfo {author} {\bibfnamefont {I.}~\bibnamefont {Esterlis}}\ and\ \bibinfo {author} {\bibfnamefont {J.}~\bibnamefont {Schmalian}},\ }\bibfield  {title} {\bibinfo {title} {Quantum critical {Eliashberg} theory},\ }\href@noop {} {\bibfield  {journal} {\bibinfo  {journal} {Annual Review of Condensed Matter Physics}\ }\textbf {\bibinfo {volume} {17}},\ \bibinfo {pages} {419} (\bibinfo {year} {2026})}\BibitemShut {NoStop}%
\bibitem [{\citenamefont {Chowdhury}\ and\ \citenamefont {Berg}(2020)}]{chowdhury2020unreasonable}%
  \BibitemOpen
  \bibfield  {author} {\bibinfo {author} {\bibfnamefont {D.}~\bibnamefont {Chowdhury}}\ and\ \bibinfo {author} {\bibfnamefont {E.}~\bibnamefont {Berg}},\ }\bibfield  {title} {\bibinfo {title} {The unreasonable effectiveness of {Eliashberg} theory for pairing of non-{Fermi} liquids},\ }\href@noop {} {\bibfield  {journal} {\bibinfo  {journal} {Annals of Physics}\ }\textbf {\bibinfo {volume} {417}},\ \bibinfo {pages} {168125} (\bibinfo {year} {2020})}\BibitemShut {NoStop}%
\bibitem [{\citenamefont {Inkof}\ \emph {et~al.}(2022)\citenamefont {Inkof}, \citenamefont {Schalm},\ and\ \citenamefont {Schmalian}}]{inkof2022quantum}%
  \BibitemOpen
  \bibfield  {author} {\bibinfo {author} {\bibfnamefont {G.-A.}\ \bibnamefont {Inkof}}, \bibinfo {author} {\bibfnamefont {K.}~\bibnamefont {Schalm}},\ and\ \bibinfo {author} {\bibfnamefont {J.}~\bibnamefont {Schmalian}},\ }\bibfield  {title} {\bibinfo {title} {Quantum critical {Eliashberg} theory, the {Sachdev-Ye-Kitaev} superconductor and their holographic duals},\ }\href@noop {} {\bibfield  {journal} {\bibinfo  {journal} {npj Quantum Materials}\ }\textbf {\bibinfo {volume} {7}},\ \bibinfo {pages} {56} (\bibinfo {year} {2022})}\BibitemShut {NoStop}%
\bibitem [{\citenamefont {Tulipman}\ \emph {et~al.}(2024)\citenamefont {Tulipman}, \citenamefont {Bashan}, \citenamefont {Schmalian},\ and\ \citenamefont {Berg}}]{tulipman2024solvable}%
  \BibitemOpen
  \bibfield  {author} {\bibinfo {author} {\bibfnamefont {E.}~\bibnamefont {Tulipman}}, \bibinfo {author} {\bibfnamefont {N.}~\bibnamefont {Bashan}}, \bibinfo {author} {\bibfnamefont {J.}~\bibnamefont {Schmalian}},\ and\ \bibinfo {author} {\bibfnamefont {E.}~\bibnamefont {Berg}},\ }\bibfield  {title} {\bibinfo {title} {Solvable models of two-level systems coupled to itinerant electrons: Robust non-{Fermi} liquid and quantum critical pairing},\ }\href@noop {} {\bibfield  {journal} {\bibinfo  {journal} {Physical Review B}\ }\textbf {\bibinfo {volume} {110}},\ \bibinfo {pages} {155118} (\bibinfo {year} {2024})}\BibitemShut {NoStop}%
\bibitem [{\citenamefont {Esterlis}\ \emph {et~al.}(2021)\citenamefont {Esterlis}, \citenamefont {Guo}, \citenamefont {Patel},\ and\ \citenamefont {Sachdev}}]{esterlis2021large}%
  \BibitemOpen
  \bibfield  {author} {\bibinfo {author} {\bibfnamefont {I.}~\bibnamefont {Esterlis}}, \bibinfo {author} {\bibfnamefont {H.}~\bibnamefont {Guo}}, \bibinfo {author} {\bibfnamefont {A.~A.}\ \bibnamefont {Patel}},\ and\ \bibinfo {author} {\bibfnamefont {S.}~\bibnamefont {Sachdev}},\ }\bibfield  {title} {\bibinfo {title} {Large-{$N$} theory of critical {Fermi} surfaces},\ }\href@noop {} {\bibfield  {journal} {\bibinfo  {journal} {Physical Review B}\ }\textbf {\bibinfo {volume} {103}},\ \bibinfo {pages} {235129} (\bibinfo {year} {2021})}\BibitemShut {NoStop}%
\bibitem [{\citenamefont {Kiessling}\ \emph {et~al.}(2025{\natexlab{c}})\citenamefont {Kiessling}, \citenamefont {Altshuler},\ and\ \citenamefont {Yuzbashyan}}]{kiessling2025bounds1}%
  \BibitemOpen
  \bibfield  {author} {\bibinfo {author} {\bibfnamefont {M.~K.-H.}\ \bibnamefont {Kiessling}}, \bibinfo {author} {\bibfnamefont {B.~L.}\ \bibnamefont {Altshuler}},\ and\ \bibinfo {author} {\bibfnamefont {E.~A.}\ \bibnamefont {Yuzbashyan}},\ }\bibfield  {title} {\bibinfo {title} {Bounds on {$T_c$} in the {Eliashberg} theory of superconductivity. {I}: The $\gamma$-model},\ }\href@noop {} {\bibfield  {journal} {\bibinfo  {journal} {Journal of Statistical Physics}\ }\textbf {\bibinfo {volume} {192}},\ \bibinfo {pages} {69} (\bibinfo {year} {2025}{\natexlab{c}})}\BibitemShut {NoStop}%
\bibitem [{\citenamefont {Allen}\ and\ \citenamefont {Dynes}(1975)}]{allen1975transition}%
  \BibitemOpen
  \bibfield  {author} {\bibinfo {author} {\bibfnamefont {P.~B.}\ \bibnamefont {Allen}}\ and\ \bibinfo {author} {\bibfnamefont {R.~C.}\ \bibnamefont {Dynes}},\ }\bibfield  {title} {\bibinfo {title} {Transition temperature of strong-coupled superconductors reanalyzed},\ }\href@noop {} {\bibfield  {journal} {\bibinfo  {journal} {Physical Review B}\ }\textbf {\bibinfo {volume} {12}},\ \bibinfo {pages} {905} (\bibinfo {year} {1975})}\BibitemShut {NoStop}%
\bibitem [{\citenamefont {Moussa}\ and\ \citenamefont {Cohen}(2006)}]{moussa2006two}%
  \BibitemOpen
  \bibfield  {author} {\bibinfo {author} {\bibfnamefont {J.~E.}\ \bibnamefont {Moussa}}\ and\ \bibinfo {author} {\bibfnamefont {M.~L.}\ \bibnamefont {Cohen}},\ }\bibfield  {title} {\bibinfo {title} {Two bounds on the maximum phonon-mediated superconducting transition temperature},\ }\href@noop {} {\bibfield  {journal} {\bibinfo  {journal} {Physical Review B—Condensed Matter and Materials Physics}\ }\textbf {\bibinfo {volume} {74}},\ \bibinfo {pages} {094520} (\bibinfo {year} {2006})}\BibitemShut {NoStop}%
\bibitem [{\citenamefont {Esterlis}\ \emph {et~al.}(2018{\natexlab{a}})\citenamefont {Esterlis}, \citenamefont {Kivelson},\ and\ \citenamefont {Scalapino}}]{esterlis2018bound}%
  \BibitemOpen
  \bibfield  {author} {\bibinfo {author} {\bibfnamefont {I.}~\bibnamefont {Esterlis}}, \bibinfo {author} {\bibfnamefont {S.~A.}\ \bibnamefont {Kivelson}},\ and\ \bibinfo {author} {\bibfnamefont {D.~J.}\ \bibnamefont {Scalapino}},\ }\bibfield  {title} {\bibinfo {title} {A bound on the superconducting transition temperature},\ }\href@noop {} {\bibfield  {journal} {\bibinfo  {journal} {npj Quantum Materials}\ }\textbf {\bibinfo {volume} {3}},\ \bibinfo {pages} {59} (\bibinfo {year} {2018}{\natexlab{a}})}\BibitemShut {NoStop}%
\bibitem [{\citenamefont {Cohen}\ and\ \citenamefont {Anderson}(1972)}]{CohenAnderson1972}%
  \BibitemOpen
  \bibfield  {author} {\bibinfo {author} {\bibfnamefont {M.~L.}\ \bibnamefont {Cohen}}\ and\ \bibinfo {author} {\bibfnamefont {P.~W.}\ \bibnamefont {Anderson}},\ }\bibfield  {title} {\bibinfo {title} {Comments on the maximum superconducting transition temperature},\ }\href {https://doi.org/10.1063/1.2946175} {\bibfield  {journal} {\bibinfo  {journal} {AIP Conference Proceedings}\ }\textbf {\bibinfo {volume} {4}},\ \bibinfo {pages} {17} (\bibinfo {year} {1972})}\BibitemShut {NoStop}%
\bibitem [{\citenamefont {Fröhlich}(1954)}]{Frohlich01071954}%
  \BibitemOpen
  \bibfield  {author} {\bibinfo {author} {\bibfnamefont {H.}~\bibnamefont {Fröhlich}},\ }\bibfield  {title} {\bibinfo {title} {Electrons in lattice fields},\ }\href {https://doi.org/10.1080/00018735400101213} {\bibfield  {journal} {\bibinfo  {journal} {Advances in Physics}\ }\textbf {\bibinfo {volume} {3}},\ \bibinfo {pages} {325} (\bibinfo {year} {1954})}\BibitemShut {NoStop}%
\bibitem [{\citenamefont {Holstein}(1959{\natexlab{a}})}]{holstein1959studies1}%
  \BibitemOpen
  \bibfield  {author} {\bibinfo {author} {\bibfnamefont {T.}~\bibnamefont {Holstein}},\ }\bibfield  {title} {\bibinfo {title} {Studies of polaron motion: Part {I}. the molecular-crystal model},\ }\href@noop {} {\bibfield  {journal} {\bibinfo  {journal} {Annals of Physics}\ }\textbf {\bibinfo {volume} {8}},\ \bibinfo {pages} {325} (\bibinfo {year} {1959}{\natexlab{a}})}\BibitemShut {NoStop}%
\bibitem [{\citenamefont {Holstein}(1959{\natexlab{b}})}]{holstein1959studies2}%
  \BibitemOpen
  \bibfield  {author} {\bibinfo {author} {\bibfnamefont {T.}~\bibnamefont {Holstein}},\ }\bibfield  {title} {\bibinfo {title} {Studies of polaron motion: Part {II}. the ``small'' polaron},\ }\href@noop {} {\bibfield  {journal} {\bibinfo  {journal} {Annals of Physics}\ }\textbf {\bibinfo {volume} {8}},\ \bibinfo {pages} {343} (\bibinfo {year} {1959}{\natexlab{b}})}\BibitemShut {NoStop}%
\bibitem [{\citenamefont {Nosarzewski}\ \emph {et~al.}(2021)\citenamefont {Nosarzewski}, \citenamefont {Huang}, \citenamefont {Dee}, \citenamefont {Esterlis}, \citenamefont {Moritz}, \citenamefont {Kivelson}, \citenamefont {Johnston},\ and\ \citenamefont {Devereaux}}]{nosarzewski2021superconductivity}%
  \BibitemOpen
  \bibfield  {author} {\bibinfo {author} {\bibfnamefont {B.}~\bibnamefont {Nosarzewski}}, \bibinfo {author} {\bibfnamefont {E.~W.}\ \bibnamefont {Huang}}, \bibinfo {author} {\bibfnamefont {P.~M.}\ \bibnamefont {Dee}}, \bibinfo {author} {\bibfnamefont {I.}~\bibnamefont {Esterlis}}, \bibinfo {author} {\bibfnamefont {B.}~\bibnamefont {Moritz}}, \bibinfo {author} {\bibfnamefont {S.~A.}\ \bibnamefont {Kivelson}}, \bibinfo {author} {\bibfnamefont {S.}~\bibnamefont {Johnston}},\ and\ \bibinfo {author} {\bibfnamefont {T.~P.}\ \bibnamefont {Devereaux}},\ }\bibfield  {title} {\bibinfo {title} {Superconductivity, charge density waves, and bipolarons in the {Holstein} model},\ }\href@noop {} {\bibfield  {journal} {\bibinfo  {journal} {Physical Review B}\ }\textbf {\bibinfo {volume} {103}},\ \bibinfo {pages} {235156} (\bibinfo {year} {2021})}\BibitemShut {NoStop}%
\bibitem [{\citenamefont {Chubukov}\ \emph {et~al.}(2026)\citenamefont {Chubukov}, \citenamefont {Esterlis}, \citenamefont {Abanov},\ and\ \citenamefont {Prokof'ev}}]{chubukov2026breakdown}%
  \BibitemOpen
  \bibfield  {author} {\bibinfo {author} {\bibfnamefont {A.}~\bibnamefont {Chubukov}}, \bibinfo {author} {\bibfnamefont {I.}~\bibnamefont {Esterlis}}, \bibinfo {author} {\bibfnamefont {A.}~\bibnamefont {Abanov}},\ and\ \bibinfo {author} {\bibfnamefont {N.}~\bibnamefont {Prokof'ev}},\ }\bibfield  {title} {\bibinfo {title} {Breakdown of the {Migdal-Eliashberg} theory for electron-phonon systems: Role of polarons and bipolarons},\ }\href {https://doi.org/10.1103/vcp5-qdfk} {\bibfield  {journal} {\bibinfo  {journal} {Physical Review B}\ }\textbf {\bibinfo {volume} {114}},\ \bibinfo {pages} {185127} (\bibinfo {year} {2026})}\BibitemShut {NoStop}%
\bibitem [{\citenamefont {Prokof'ev}\ \emph {et~al.}(2026)\citenamefont {Prokof'ev}, \citenamefont {Esterlis}, \citenamefont {Abanov},\ and\ \citenamefont {Chubukov}}]{prokof2026limits}%
  \BibitemOpen
  \bibfield  {author} {\bibinfo {author} {\bibfnamefont {N.}~\bibnamefont {Prokof'ev}}, \bibinfo {author} {\bibfnamefont {I.}~\bibnamefont {Esterlis}}, \bibinfo {author} {\bibfnamefont {A.}~\bibnamefont {Abanov}},\ and\ \bibinfo {author} {\bibfnamefont {A.}~\bibnamefont {Chubukov}},\ }\bibfield  {title} {\bibinfo {title} {Limits of validity for {Migdal-Eliashberg} theory: Role of polarons and bipolarons},\ }\href {https://doi.org/10.1103/zd4s-cgnj} {\bibfield  {journal} {\bibinfo  {journal} {Physical Review Letters}\ }\textbf {\bibinfo {volume} {137}},\ \bibinfo {pages} {096501} (\bibinfo {year} {2026})}\BibitemShut {NoStop}%
\bibitem [{\citenamefont {Mishchenko}\ \emph {et~al.}(2000)\citenamefont {Mishchenko}, \citenamefont {Prokof'ev}, \citenamefont {Sakamoto},\ and\ \citenamefont {Svistunov}}]{mishchenko2000diagrammatic}%
  \BibitemOpen
  \bibfield  {author} {\bibinfo {author} {\bibfnamefont {A.~S.}\ \bibnamefont {Mishchenko}}, \bibinfo {author} {\bibfnamefont {N.~V.}\ \bibnamefont {Prokof'ev}}, \bibinfo {author} {\bibfnamefont {A.}~\bibnamefont {Sakamoto}},\ and\ \bibinfo {author} {\bibfnamefont {B.~V.}\ \bibnamefont {Svistunov}},\ }\bibfield  {title} {\bibinfo {title} {Diagrammatic quantum {Monte Carlo} study of the {Fr{\"o}hlich} polaron},\ }\href@noop {} {\bibfield  {journal} {\bibinfo  {journal} {Physical Review B}\ }\textbf {\bibinfo {volume} {62}},\ \bibinfo {pages} {6317} (\bibinfo {year} {2000})}\BibitemShut {NoStop}%
\bibitem [{\citenamefont {Prokof'ev}\ and\ \citenamefont {Svistunov}(1998)}]{prokof1998polaron}%
  \BibitemOpen
  \bibfield  {author} {\bibinfo {author} {\bibfnamefont {N.~V.}\ \bibnamefont {Prokof'ev}}\ and\ \bibinfo {author} {\bibfnamefont {B.~V.}\ \bibnamefont {Svistunov}},\ }\bibfield  {title} {\bibinfo {title} {Polaron problem by diagrammatic quantum {Monte Carlo}},\ }\href@noop {} {\bibfield  {journal} {\bibinfo  {journal} {Physical Review Letters}\ }\textbf {\bibinfo {volume} {81}},\ \bibinfo {pages} {2514} (\bibinfo {year} {1998})}\BibitemShut {NoStop}%
\bibitem [{\citenamefont {Esterlis}\ \emph {et~al.}(2018{\natexlab{b}})\citenamefont {Esterlis}, \citenamefont {Nosarzewski}, \citenamefont {Huang}, \citenamefont {Moritz}, \citenamefont {Devereaux}, \citenamefont {Scalapino},\ and\ \citenamefont {Kivelson}}]{esterlis2018breakdown}%
  \BibitemOpen
  \bibfield  {author} {\bibinfo {author} {\bibfnamefont {I.}~\bibnamefont {Esterlis}}, \bibinfo {author} {\bibfnamefont {B.}~\bibnamefont {Nosarzewski}}, \bibinfo {author} {\bibfnamefont {E.~W.}\ \bibnamefont {Huang}}, \bibinfo {author} {\bibfnamefont {B.}~\bibnamefont {Moritz}}, \bibinfo {author} {\bibfnamefont {T.~P.}\ \bibnamefont {Devereaux}}, \bibinfo {author} {\bibfnamefont {D.~J.}\ \bibnamefont {Scalapino}},\ and\ \bibinfo {author} {\bibfnamefont {S.~A.}\ \bibnamefont {Kivelson}},\ }\bibfield  {title} {\bibinfo {title} {Breakdown of the {Migdal-Eliashberg} theory: A determinant quantum {Monte Carlo} study},\ }\href@noop {} {\bibfield  {journal} {\bibinfo  {journal} {Physical Review B}\ }\textbf {\bibinfo {volume} {97}},\ \bibinfo {pages} {140501} (\bibinfo {year} {2018}{\natexlab{b}})}\BibitemShut {NoStop}%
\bibitem [{\citenamefont {Fitzpatrick}\ \emph {et~al.}(2014)\citenamefont {Fitzpatrick}, \citenamefont {Kachru}, \citenamefont {Kaplan},\ and\ \citenamefont {Raghu}}]{fitzpatrick2014non}%
  \BibitemOpen
  \bibfield  {author} {\bibinfo {author} {\bibfnamefont {A.~L.}\ \bibnamefont {Fitzpatrick}}, \bibinfo {author} {\bibfnamefont {S.}~\bibnamefont {Kachru}}, \bibinfo {author} {\bibfnamefont {J.}~\bibnamefont {Kaplan}},\ and\ \bibinfo {author} {\bibfnamefont {S.}~\bibnamefont {Raghu}},\ }\bibfield  {title} {\bibinfo {title} {Non-{Fermi}-liquid behavior of large-{$N_B$} quantum critical metals},\ }\href@noop {} {\bibfield  {journal} {\bibinfo  {journal} {Physical Review B}\ }\textbf {\bibinfo {volume} {89}},\ \bibinfo {pages} {165114} (\bibinfo {year} {2014})}\BibitemShut {NoStop}%
\bibitem [{\citenamefont {Werman}\ and\ \citenamefont {Berg}(2016)}]{Werman2016}%
  \BibitemOpen
  \bibfield  {author} {\bibinfo {author} {\bibfnamefont {Y.}~\bibnamefont {Werman}}\ and\ \bibinfo {author} {\bibfnamefont {E.}~\bibnamefont {Berg}},\ }\bibfield  {title} {\bibinfo {title} {{Mott-Ioffe-Regel} limit and resistivity crossover in a tractable electron-phonon model},\ }\href {https://doi.org/10.1103/PhysRevB.93.075109} {\bibfield  {journal} {\bibinfo  {journal} {Phys. Rev. B}\ }\textbf {\bibinfo {volume} {93}},\ \bibinfo {pages} {075109} (\bibinfo {year} {2016})}\BibitemShut {NoStop}%
\bibitem [{\citenamefont {Werman}\ \emph {et~al.}(2017)\citenamefont {Werman}, \citenamefont {Kivelson},\ and\ \citenamefont {Berg}}]{werman2017non}%
  \BibitemOpen
  \bibfield  {author} {\bibinfo {author} {\bibfnamefont {Y.}~\bibnamefont {Werman}}, \bibinfo {author} {\bibfnamefont {S.~A.}\ \bibnamefont {Kivelson}},\ and\ \bibinfo {author} {\bibfnamefont {E.}~\bibnamefont {Berg}},\ }\bibfield  {title} {\bibinfo {title} {Non-quasiparticle transport and resistivity saturation: a view from the large-{$N$} limit},\ }\href@noop {} {\bibfield  {journal} {\bibinfo  {journal} {npj Quantum Materials}\ }\textbf {\bibinfo {volume} {2}},\ \bibinfo {pages} {7} (\bibinfo {year} {2017})}\BibitemShut {NoStop}%
\bibitem [{\citenamefont {Hofmann}\ \emph {et~al.}(2022)\citenamefont {Hofmann}, \citenamefont {Chowdhury}, \citenamefont {Kivelson},\ and\ \citenamefont {Berg}}]{hofmann2022heuristic}%
  \BibitemOpen
  \bibfield  {author} {\bibinfo {author} {\bibfnamefont {J.~S.}\ \bibnamefont {Hofmann}}, \bibinfo {author} {\bibfnamefont {D.}~\bibnamefont {Chowdhury}}, \bibinfo {author} {\bibfnamefont {S.~A.}\ \bibnamefont {Kivelson}},\ and\ \bibinfo {author} {\bibfnamefont {E.}~\bibnamefont {Berg}},\ }\bibfield  {title} {\bibinfo {title} {Heuristic bounds on superconductivity and how to exceed them},\ }\href@noop {} {\bibfield  {journal} {\bibinfo  {journal} {npj Quantum Materials}\ }\textbf {\bibinfo {volume} {7}},\ \bibinfo {pages} {83} (\bibinfo {year} {2022})}\BibitemShut {NoStop}%
\bibitem [{\citenamefont {Yuzbashyan}\ and\ \citenamefont {Altshuler}(2022{\natexlab{b}})}]{yuzbashyan2022breakdown}%
  \BibitemOpen
  \bibfield  {author} {\bibinfo {author} {\bibfnamefont {E.~A.}\ \bibnamefont {Yuzbashyan}}\ and\ \bibinfo {author} {\bibfnamefont {B.~L.}\ \bibnamefont {Altshuler}},\ }\bibfield  {title} {\bibinfo {title} {Breakdown of the {Migdal-Eliashberg} theory and a theory of lattice-fermionic superfluidity},\ }\href@noop {} {\bibfield  {journal} {\bibinfo  {journal} {Physical Review B}\ }\textbf {\bibinfo {volume} {106}},\ \bibinfo {pages} {054518} (\bibinfo {year} {2022}{\natexlab{b}})}\BibitemShut {NoStop}%
\bibitem [{\citenamefont {Zhang}\ \emph {et~al.}(2023)\citenamefont {Zhang}, \citenamefont {Berg},\ and\ \citenamefont {Chubukov}}]{zhang2023free}%
  \BibitemOpen
  \bibfield  {author} {\bibinfo {author} {\bibfnamefont {S.-S.}\ \bibnamefont {Zhang}}, \bibinfo {author} {\bibfnamefont {E.}~\bibnamefont {Berg}},\ and\ \bibinfo {author} {\bibfnamefont {A.~V.}\ \bibnamefont {Chubukov}},\ }\bibfield  {title} {\bibinfo {title} {Free energy and specific heat near a quantum critical point of a metal},\ }\href@noop {} {\bibfield  {journal} {\bibinfo  {journal} {Physical Review B}\ }\textbf {\bibinfo {volume} {107}},\ \bibinfo {pages} {144507} (\bibinfo {year} {2023})}\BibitemShut {NoStop}%
\bibitem [{\citenamefont {Yuzbashyan}\ \emph {et~al.}(2025)\citenamefont {Yuzbashyan}, \citenamefont {Altshuler},\ and\ \citenamefont {Patra}}]{yuzbashyan2025instability}%
  \BibitemOpen
  \bibfield  {author} {\bibinfo {author} {\bibfnamefont {E.~A.}\ \bibnamefont {Yuzbashyan}}, \bibinfo {author} {\bibfnamefont {B.~L.}\ \bibnamefont {Altshuler}},\ and\ \bibinfo {author} {\bibfnamefont {A.}~\bibnamefont {Patra}},\ }\bibfield  {title} {\bibinfo {title} {Instability of metals with respect to strong electron-phonon interaction},\ }\href@noop {} {\bibfield  {journal} {\bibinfo  {journal} {Physical Review Letters}\ }\textbf {\bibinfo {volume} {135}},\ \bibinfo {pages} {026503} (\bibinfo {year} {2025})}\BibitemShut {NoStop}%
\bibitem [{\citenamefont {Semenok}\ \emph {et~al.}(2025)\citenamefont {Semenok}, \citenamefont {Altshuler},\ and\ \citenamefont {Yuzbashyan}}]{semenok2025}%
  \BibitemOpen
  \bibfield  {author} {\bibinfo {author} {\bibfnamefont {D.~V.}\ \bibnamefont {Semenok}}, \bibinfo {author} {\bibfnamefont {B.~L.}\ \bibnamefont {Altshuler}},\ and\ \bibinfo {author} {\bibfnamefont {E.~A.}\ \bibnamefont {Yuzbashyan}},\ }\bibfield  {title} {\bibinfo {title} {Fundamental limits on the electron-phonon coupling and superconducting {$T_c$}},\ }\href {https://doi.org/10.1002/adma.202507013} {\bibfield  {journal} {\bibinfo  {journal} {Advanced Materials}\ }\textbf {\bibinfo {volume} {37}},\ \bibinfo {pages} {2507013} (\bibinfo {year} {2025})}\BibitemShut {NoStop}%
\bibitem [{\citenamefont {Abrikosov}\ \emph {et~al.}(1963)\citenamefont {Abrikosov}, \citenamefont {Gorkov},\ and\ \citenamefont {Dzyaloshinski}}]{abrikosov1964methods}%
  \BibitemOpen
  \bibfield  {author} {\bibinfo {author} {\bibfnamefont {A.~A.}\ \bibnamefont {Abrikosov}}, \bibinfo {author} {\bibfnamefont {L.~P.}\ \bibnamefont {Gorkov}},\ and\ \bibinfo {author} {\bibfnamefont {I.~E.}\ \bibnamefont {Dzyaloshinski}},\ }\href@noop {} {\emph {\bibinfo {title} {Methods of quantum field theory in statistical physics}}}\ (\bibinfo  {publisher} {Prentice-Hall},\ \bibinfo {year} {1963})\ \bibinfo {note} {translated and edited by Richard A. Silverman}\BibitemShut {NoStop}%
\bibitem [{\citenamefont {Prange}\ and\ \citenamefont {Kadanoff}(1964)}]{PrangeKadanoff}%
  \BibitemOpen
  \bibfield  {author} {\bibinfo {author} {\bibfnamefont {R.~E.}\ \bibnamefont {Prange}}\ and\ \bibinfo {author} {\bibfnamefont {L.~P.}\ \bibnamefont {Kadanoff}},\ }\bibfield  {title} {\bibinfo {title} {Transport theory for electron-phonon interactions in metals},\ }\href {https://doi.org/10.1103/PhysRev.134.A566} {\bibfield  {journal} {\bibinfo  {journal} {Phys. Rev.}\ }\textbf {\bibinfo {volume} {134}},\ \bibinfo {pages} {A566} (\bibinfo {year} {1964})}\BibitemShut {NoStop}%
\bibitem [{\citenamefont {Rammer}\ and\ \citenamefont {Smith}(1986)}]{rammer1986quantum}%
  \BibitemOpen
  \bibfield  {author} {\bibinfo {author} {\bibfnamefont {J.}~\bibnamefont {Rammer}}\ and\ \bibinfo {author} {\bibfnamefont {H.}~\bibnamefont {Smith}},\ }\bibfield  {title} {\bibinfo {title} {Quantum field-theoretical methods in transport theory of metals},\ }\href@noop {} {\bibfield  {journal} {\bibinfo  {journal} {Reviews of Modern Physics}\ }\textbf {\bibinfo {volume} {58}},\ \bibinfo {pages} {323} (\bibinfo {year} {1986})}\BibitemShut {NoStop}%
\bibitem [{Note1()}]{Note1}%
  \BibitemOpen
  \bibinfo {note} {An extreme example of this is the non-causal nature of the Abraham-Lorentz force \cite {efimkin2016non}, whose Markovian approximation is non-causal even at arbitrarily weak couplings.}\BibitemShut {Stop}%
\bibitem [{Note2()}]{Note2}%
  \BibitemOpen
  \bibinfo {note} {Since $\partial _\gamma \protect \hat {C}_{\gamma =0}$ generally has negative eigenvalues, solutions to Eq.~\protect \eqref {lin QBE} can always be found at large enough values of $|\gamma |$, where it approaches $\protect \hat {K}_\gamma \approx \gamma \left (1 + \protect \frac {g^2}{T}\partial _\gamma \protect \hat {C}_{\gamma =0}\right )$. Thus the ``critical'' coupling predicted by the QBE is $g_{c}^2/T= 1/\max \left |{\protect \rm eig}\left (\partial _\gamma \protect \hat {C}_{\gamma =0}\right )\right |$. Interestingly, this eigenvalue diverges in the $\Omega _0/T\to 0$ limit.}\BibitemShut {Stop}%
\bibitem [{Note3()}]{Note3}%
  \BibitemOpen
  \bibinfo {note} {This method is equivalent to a calculation of the response function using the Kubo formula, including Bethe-Salpeter re-summation of all ``ladder diagrams''.}\BibitemShut {Stop}%
\bibitem [{Note4()}]{Note4}%
  \BibitemOpen
  \bibinfo {note} {We note that these parameters, corresponding to $\lambda =6,T/g=0.41$ would appear to be in the QBE's stable regime according to Table~1 in the supplementary material of \cite {yuzbashyan2025instability}. We believe this is due to a missing factor of $1/dE^2$ in the discretization of the principal-value term of Eq.~22 of their supplementary material.}\BibitemShut {Stop}%
\bibitem [{\citenamefont {Altland}\ and\ \citenamefont {Simons}(2010)}]{AltlandSimons2010}%
  \BibitemOpen
  \bibfield  {author} {\bibinfo {author} {\bibfnamefont {A.}~\bibnamefont {Altland}}\ and\ \bibinfo {author} {\bibfnamefont {B.~D.}\ \bibnamefont {Simons}},\ }\href@noop {} {\emph {\bibinfo {title} {Condensed Matter Field Theory}}},\ \bibinfo {edition} {2nd}\ ed.\ (\bibinfo  {publisher} {Cambridge University Press},\ \bibinfo {year} {2010})\BibitemShut {NoStop}%
\bibitem [{Note5()}]{Note5}%
  \BibitemOpen
  \bibinfo {note} {Note that we consider relaxation solely in the presence of strong el-ph coupling. A common yet different analysis is of relaxation in the presence of el-el interactions which are significantly stronger than the el-ph interactions, such that the bottleneck for relaxation is the perturbative transfer of energy between electrons and phonons \cite {allen1987theory,SciPostPhys.13.4.095,guo2026phonon}. In that case, at timescales longer than the el-el relaxation time, perturbations to the electrons can be projected to a family of thermal distributions with time dependence carried by the electron's temperature. This agrees with our results at sufficiently small el-ph coupling, as shown in Appendix~\ref {app: relaxation}. We note that it is not clear a-priori that a similar separation of timescales exists in the case where both el-el and el-ep interactions are sufficiently strong; as we show in the rest of this work, stronger interactions do not necessarily lead to faster relaxation, and the
  relaxation rate may instead saturate at a ``Planckian'' limit.}\BibitemShut {Stop}%
\bibitem [{\citenamefont {Chen}\ \emph {et~al.}(1971)\citenamefont {Chen}, \citenamefont {Leslie},\ and\ \citenamefont {Smith}}]{chen1971electron}%
  \BibitemOpen
  \bibfield  {author} {\bibinfo {author} {\bibfnamefont {T.~T.}\ \bibnamefont {Chen}}, \bibinfo {author} {\bibfnamefont {J.~D.}\ \bibnamefont {Leslie}},\ and\ \bibinfo {author} {\bibfnamefont {H.~J.~T.}\ \bibnamefont {Smith}},\ }\bibfield  {title} {\bibinfo {title} {Electron tunneling study of amorphous {Pb-Bi} superconducting alloys},\ }\href@noop {} {\bibfield  {journal} {\bibinfo  {journal} {Physica}\ }\textbf {\bibinfo {volume} {55}},\ \bibinfo {pages} {439} (\bibinfo {year} {1971})}\BibitemShut {NoStop}%
\bibitem [{\citenamefont {Semenok}\ \emph {et~al.}(2021)\citenamefont {Semenok}, \citenamefont {Troyan}, \citenamefont {Ivanova}, \citenamefont {Kvashnin}, \citenamefont {Kruglov}, \citenamefont {Hanfland}, \citenamefont {Sadakov}, \citenamefont {Sobolevskiy}, \citenamefont {Pervakov}, \citenamefont {Lyubutin} \emph {et~al.}}]{semenok2021superconductivity}%
  \BibitemOpen
  \bibfield  {author} {\bibinfo {author} {\bibfnamefont {D.~V.}\ \bibnamefont {Semenok}}, \bibinfo {author} {\bibfnamefont {I.~A.}\ \bibnamefont {Troyan}}, \bibinfo {author} {\bibfnamefont {A.~G.}\ \bibnamefont {Ivanova}}, \bibinfo {author} {\bibfnamefont {A.~G.}\ \bibnamefont {Kvashnin}}, \bibinfo {author} {\bibfnamefont {I.~A.}\ \bibnamefont {Kruglov}}, \bibinfo {author} {\bibfnamefont {M.}~\bibnamefont {Hanfland}}, \bibinfo {author} {\bibfnamefont {A.~V.}\ \bibnamefont {Sadakov}}, \bibinfo {author} {\bibfnamefont {O.~A.}\ \bibnamefont {Sobolevskiy}}, \bibinfo {author} {\bibfnamefont {K.~S.}\ \bibnamefont {Pervakov}}, \bibinfo {author} {\bibfnamefont {I.~S.}\ \bibnamefont {Lyubutin}}, \emph {et~al.},\ }\bibfield  {title} {\bibinfo {title} {Superconductivity at {253 K} in lanthanum--yttrium ternary hydrides},\ }\href@noop {} {\bibfield  {journal} {\bibinfo  {journal} {Materials Today}\ }\textbf {\bibinfo {volume} {48}},\ \bibinfo {pages} {18} (\bibinfo {year} {2021})}\BibitemShut {NoStop}%
\bibitem [{\citenamefont {Kumar}\ \emph {et~al.}(2025)\citenamefont {Kumar}, \citenamefont {Barbalas}, \citenamefont {Bhandia}, \citenamefont {Lee}, \citenamefont {Varshney}, \citenamefont {Jalan},\ and\ \citenamefont {Armitage}}]{kumar2025absencetwophononquasielasticscattering}%
  \BibitemOpen
  \bibfield  {author} {\bibinfo {author} {\bibfnamefont {K.~S.}\ \bibnamefont {Kumar}}, \bibinfo {author} {\bibfnamefont {D.}~\bibnamefont {Barbalas}}, \bibinfo {author} {\bibfnamefont {R.}~\bibnamefont {Bhandia}}, \bibinfo {author} {\bibfnamefont {D.}~\bibnamefont {Lee}}, \bibinfo {author} {\bibfnamefont {S.}~\bibnamefont {Varshney}}, \bibinfo {author} {\bibfnamefont {B.}~\bibnamefont {Jalan}},\ and\ \bibinfo {author} {\bibfnamefont {N.~P.}\ \bibnamefont {Armitage}},\ }\href {https://arxiv.org/abs/2501.15771} {\bibinfo {title} {Absence of two-phonon quasi-elastic scattering in the normal state of doped--{SrTiO$_3$} by {THz} pump-probe spectroscopy}} (\bibinfo {year} {2025}),\ \Eprint {https://arxiv.org/abs/2501.15771} {arXiv:2501.15771 [cond-mat.str-el]} \BibitemShut {NoStop}%
\bibitem [{\citenamefont {Barbalas}\ \emph {et~al.}(2025)\citenamefont {Barbalas}, \citenamefont {Romero}, \citenamefont {Chaudhuri}, \citenamefont {Mahmood}, \citenamefont {Nair}, \citenamefont {Schreiber}, \citenamefont {Schlom}, \citenamefont {Shen},\ and\ \citenamefont {Armitage}}]{barbalas2025energy}%
  \BibitemOpen
  \bibfield  {author} {\bibinfo {author} {\bibfnamefont {D.}~\bibnamefont {Barbalas}}, \bibinfo {author} {\bibfnamefont {R.}~\bibnamefont {Romero}, \bibfnamefont {III}}, \bibinfo {author} {\bibfnamefont {D.}~\bibnamefont {Chaudhuri}}, \bibinfo {author} {\bibfnamefont {F.}~\bibnamefont {Mahmood}}, \bibinfo {author} {\bibfnamefont {H.~P.}\ \bibnamefont {Nair}}, \bibinfo {author} {\bibfnamefont {N.~J.}\ \bibnamefont {Schreiber}}, \bibinfo {author} {\bibfnamefont {D.~G.}\ \bibnamefont {Schlom}}, \bibinfo {author} {\bibfnamefont {K.~M.}\ \bibnamefont {Shen}},\ and\ \bibinfo {author} {\bibfnamefont {N.~P.}\ \bibnamefont {Armitage}},\ }\bibfield  {title} {\bibinfo {title} {Energy relaxation and dynamics in the correlated metal {$\mathrm{Sr}_2\mathrm{RuO}_4$} via terahertz two-dimensional coherent spectroscopy},\ }\href@noop {} {\bibfield  {journal} {\bibinfo  {journal} {Physical Review Letters}\ }\textbf {\bibinfo {volume} {134}},\ \bibinfo {pages} {036501} (\bibinfo {year} {2025})}\BibitemShut {NoStop}%
\bibitem [{\citenamefont {Hartnoll}\ and\ \citenamefont {Mackenzie}(2022)}]{hartnoll2022colloquium}%
  \BibitemOpen
  \bibfield  {author} {\bibinfo {author} {\bibfnamefont {S.~A.}\ \bibnamefont {Hartnoll}}\ and\ \bibinfo {author} {\bibfnamefont {A.~P.}\ \bibnamefont {Mackenzie}},\ }\bibfield  {title} {\bibinfo {title} {Colloquium: Planckian dissipation in metals},\ }\href@noop {} {\bibfield  {journal} {\bibinfo  {journal} {Reviews of Modern Physics}\ }\textbf {\bibinfo {volume} {94}},\ \bibinfo {pages} {041002} (\bibinfo {year} {2022})}\BibitemShut {NoStop}%
\bibitem [{\citenamefont {Sachdev}(1999)}]{sachdev1999universal}%
  \BibitemOpen
  \bibfield  {author} {\bibinfo {author} {\bibfnamefont {S.}~\bibnamefont {Sachdev}},\ }\bibfield  {title} {\bibinfo {title} {Universal relaxational dynamics near two-dimensional quantum critical points},\ }\href@noop {} {\bibfield  {journal} {\bibinfo  {journal} {Physical Review B}\ }\textbf {\bibinfo {volume} {59}},\ \bibinfo {pages} {14054} (\bibinfo {year} {1999})}\BibitemShut {NoStop}%
\bibitem [{\citenamefont {Zaanen}(2004)}]{zaanen2004temperature}%
  \BibitemOpen
  \bibfield  {author} {\bibinfo {author} {\bibfnamefont {J.}~\bibnamefont {Zaanen}},\ }\bibfield  {title} {\bibinfo {title} {Why the temperature is high},\ }\href@noop {} {\bibfield  {journal} {\bibinfo  {journal} {Nature}\ }\textbf {\bibinfo {volume} {430}},\ \bibinfo {pages} {512} (\bibinfo {year} {2004})}\BibitemShut {NoStop}%
\bibitem [{\citenamefont {Qi}\ \emph {et~al.}(2026)\citenamefont {Qi}, \citenamefont {Milekhin},\ and\ \citenamefont {Delacr{\'e}taz}}]{qi2026planckian}%
  \BibitemOpen
  \bibfield  {author} {\bibinfo {author} {\bibfnamefont {M.}~\bibnamefont {Qi}}, \bibinfo {author} {\bibfnamefont {A.}~\bibnamefont {Milekhin}},\ and\ \bibinfo {author} {\bibfnamefont {L.}~\bibnamefont {Delacr{\'e}taz}},\ }\bibfield  {title} {\bibinfo {title} {Planckian bound on the local equilibration time},\ }\href@noop {} {\bibfield  {journal} {\bibinfo  {journal} {arXiv preprint arXiv:2602.17638}\ } (\bibinfo {year} {2026})}\BibitemShut {NoStop}%
\bibitem [{\citenamefont {Guo}(2024)}]{GuoFluctation}%
  \BibitemOpen
  \bibfield  {author} {\bibinfo {author} {\bibfnamefont {H.}~\bibnamefont {Guo}},\ }\bibfield  {title} {\bibinfo {title} {Fluctuation spectrum of critical {Fermi} surfaces},\ }\href {https://doi.org/10.1103/PhysRevB.110.155130} {\bibfield  {journal} {\bibinfo  {journal} {Phys. Rev. B}\ }\textbf {\bibinfo {volume} {110}},\ \bibinfo {pages} {155130} (\bibinfo {year} {2024})}\BibitemShut {NoStop}%
\bibitem [{\citenamefont {Efimkin}\ \emph {et~al.}(2016)\citenamefont {Efimkin}, \citenamefont {Hofmann},\ and\ \citenamefont {Galitski}}]{efimkin2016non}%
  \BibitemOpen
  \bibfield  {author} {\bibinfo {author} {\bibfnamefont {D.~K.}\ \bibnamefont {Efimkin}}, \bibinfo {author} {\bibfnamefont {J.}~\bibnamefont {Hofmann}},\ and\ \bibinfo {author} {\bibfnamefont {V.}~\bibnamefont {Galitski}},\ }\bibfield  {title} {\bibinfo {title} {Non-markovian quantum friction of bright solitons in superfluids},\ }\href@noop {} {\bibfield  {journal} {\bibinfo  {journal} {Physical Review Letters}\ }\textbf {\bibinfo {volume} {116}},\ \bibinfo {pages} {225301} (\bibinfo {year} {2016})}\BibitemShut {NoStop}%
\bibitem [{\citenamefont {Allen}(1987)}]{allen1987theory}%
  \BibitemOpen
  \bibfield  {author} {\bibinfo {author} {\bibfnamefont {P.~B.}\ \bibnamefont {Allen}},\ }\bibfield  {title} {\bibinfo {title} {Theory of thermal relaxation of electrons in metals},\ }\href@noop {} {\bibfield  {journal} {\bibinfo  {journal} {Physical Review Letters}\ }\textbf {\bibinfo {volume} {59}},\ \bibinfo {pages} {1460} (\bibinfo {year} {1987})}\BibitemShut {NoStop}%
\bibitem [{\citenamefont {Glorioso}\ and\ \citenamefont {Hartnoll}(2022)}]{SciPostPhys.13.4.095}%
  \BibitemOpen
  \bibfield  {author} {\bibinfo {author} {\bibfnamefont {P.}~\bibnamefont {Glorioso}}\ and\ \bibinfo {author} {\bibfnamefont {S.~A.}\ \bibnamefont {Hartnoll}},\ }\bibfield  {title} {\bibinfo {title} {Joule heating in bad and slow metals},\ }\href {https://doi.org/10.21468/SciPostPhys.13.4.095} {\bibfield  {journal} {\bibinfo  {journal} {SciPost Phys.}\ }\textbf {\bibinfo {volume} {13}},\ \bibinfo {pages} {095} (\bibinfo {year} {2022})}\BibitemShut {NoStop}%
\bibitem [{\citenamefont {Guo}\ and\ \citenamefont {Chowdhury}(2026)}]{guo2026phonon}%
  \BibitemOpen
  \bibfield  {author} {\bibinfo {author} {\bibfnamefont {H.}~\bibnamefont {Guo}}\ and\ \bibinfo {author} {\bibfnamefont {D.}~\bibnamefont {Chowdhury}},\ }\bibfield  {title} {\bibinfo {title} {Phonon induced energy relaxation in quantum critical metals},\ }\href@noop {} {\bibfield  {journal} {\bibinfo  {journal} {Physical Review Letters}\ }\textbf {\bibinfo {volume} {136}},\ \bibinfo {pages} {126503} (\bibinfo {year} {2026})}\BibitemShut {NoStop}%
\bibitem [{\citenamefont {Hale}\ and\ \citenamefont {Verduyn~Lunel}(1993)}]{hale1993functional}%
  \BibitemOpen
  \bibfield  {author} {\bibinfo {author} {\bibfnamefont {J.~K.}\ \bibnamefont {Hale}}\ and\ \bibinfo {author} {\bibfnamefont {S.~M.}\ \bibnamefont {Verduyn~Lunel}},\ }\href {https://doi.org/10.1007/978-1-4612-4342-7} {\emph {\bibinfo {title} {Introduction to Functional Differential Equations}}},\ \bibinfo {series} {Applied Mathematical Sciences}, Vol.~\bibinfo {volume} {99}\ (\bibinfo  {publisher} {Springer},\ \bibinfo {address} {New York},\ \bibinfo {year} {1993})\BibitemShut {NoStop}%
\bibitem [{Note6()}]{Note6}%
  \BibitemOpen
  \bibinfo {note} {We note that, to the best of our knowledge, there is no fundamental constraint preventing the eigenfunctions from having tails that grow exponentially with $|\tau |$. Furthermore, it is not difficult to construct examples where an RFDE has eigenfunctions with exponentially growing tails, but the solution of any IVP with $L^2$ initial conditions remains $L^2$ at any finite time. A simple example similar in spirit to the KBE is $df(t,\tau )/dt = -\intop \ilimits@ _0^\infty e^{-(\tau +s)}f(t-\tau ,s)ds$}\BibitemShut {NoStop}%
\bibitem [{\citenamefont {P{\"o}schl}\ and\ \citenamefont {Teller}(1933)}]{poschl1933bemerkungen}%
  \BibitemOpen
  \bibfield  {author} {\bibinfo {author} {\bibfnamefont {G.}~\bibnamefont {P{\"o}schl}}\ and\ \bibinfo {author} {\bibfnamefont {E.}~\bibnamefont {Teller}},\ }\bibfield  {title} {\bibinfo {title} {Bemerkungen zur quantenmechanik des anharmonischen oszillators},\ }\href@noop {} {\bibfield  {journal} {\bibinfo  {journal} {Zeitschrift f{\"u}r Physik}\ }\textbf {\bibinfo {volume} {83}},\ \bibinfo {pages} {143} (\bibinfo {year} {1933})}\BibitemShut {NoStop}%
\bibitem [{Note7()}]{Note7}%
  \BibitemOpen
  \bibinfo {note} {The slope with respect to $\gamma $ must be positive; if it were negative, the eigenvalues would have become negative at positive $\gamma $, implying an instability. This can be verified easily by a direct calculation of the matrix elements of this mode with $\partial _\gamma \protect \hat {C}_{0}$.}\BibitemShut {Stop}%
\end{thebibliography}%
\clearpage
\twocolumngrid
\section*{End Matter}

In the self-consistent one-loop approximation, the electron self-energies are given by \cite{rammer1986quantum}
\begin{equation}
\begin{aligned}
\Sigma^R(t_1,t_2)&= D^R(t_1-t_2)g^K(t_1,t_2)\\ &\quad +D^K(t_1-t_2)g^R(t_1,t_2), \\  
\Sigma^A(t_1,t_2)&= D^A(t_1-t_2)g^K(t_1,t_2)\\ &\quad +D^K(t_1-t_2)g^A(t_1,t_2), \\  
\Sigma^K(t_1,t_2)&= \bigl(D^R(t_1-t_2)-D^A(t_1-t_2)\bigr)\\ &\quad \bigl(g^R(t_1,t_2)-g^A(t_1,t_2)\bigr)\\ &\quad +D^K(t_1-t_2)g^K(t_1,t_2),
\end{aligned}
\label{self energy}
\end{equation}
where the retarded/advanced components of the electron's quasiclassical Green's functions are
\begin{equation}
    g^{R/A}(t,E)=\frac{i}{\pi}\int G^{R/A}(t,E,\xi)d\xi \approx \pm 1
\end{equation}
in the infinite bandwidth limit, having assumed a momentum-independent self energy. The phonon propagators are (including the interaction strength factors)
\begin{equation}
\begin{aligned}
D^{R/A}(E) &= \frac{\pi g^2}{4\Omega_0}\bigl(\frac{1}{E-\Omega_0 \pm i0} - \frac{1}{E+\Omega_0 \pm i0}\bigr), \\   
    \mu(E)&\equiv i(D^R(E)-D^A(E))\\ &= \frac{\pi^2 g^2}{2\Omega_0}\bigl(\delta(E-\Omega_0) - \delta(E+\Omega_0)\bigr), \\  
    iD^{K}(E) &=\coth\bigl(\frac{E}{2T}\bigr) \mu(E)\\ &= \frac{\pi}{2} \Gamma_0 \bigl(\delta(E-\Omega_0) + \delta(E+\Omega_0)\bigr).
\end{aligned}
\end{equation}
where $\Gamma_0=\frac{\pi g^2}{\Omega_0}\coth\left(\frac{\Omega_0} {2T}\right)$ is the quasielastic scattering rate.
% or as a function of time difference
% \begin{align}
%     D^{R/A}(\tau)&=\pm \frac{\pi g^2}{2\Omega_0} \Theta(\pm t)\sin(\Omega_0t)
%     \\ \nonumber D^{K}(\tau)&=\frac{\pi^2 g^2}{\Omega_0} \coth\left(\frac{\Omega_0}{2T}\right)\cos(\Omega_0t).
% \end{align}

In equilibrium, using $g^K_{\rm eq}(E)=2\tanh\left(\frac{E}{2T}\right)$ gives the equilibrium self-energies:

\begin{widetext}
\begin{align}
    \Sigma^{R/A}_{\rm eq}(E)&= \underbrace{\frac{ g^2}{2\Omega_0}\left(\psi\left(\frac{1}{2}\mp i\frac{E+\Omega_0}{2\pi T}\right)-\psi\left(\frac{1}{2}\mp i\frac{E-\Omega_0}{2\pi T}\right)\right)}_{\equiv\sigma^{R/A}(E)} \mp \frac{i}{2} \Gamma_0\\ \nonumber 
    \Gamma_{\rm eq}(E)&\equiv i\left(\Sigma^{R}_{\rm eq}(E)-\Sigma^{A}_{\rm eq}(E)\right) = \Gamma_0 -  \frac{\pi g^2}{2\Omega_0} \left(\tanh\left(\frac{E+\Omega_0}{2T}\right)-\tanh\left(\frac{E-\Omega_0}{2T}\right)\right) 
\end{align}
\end{widetext}
where $\psi(z)$ is the digamma function. The equilibrium Keldysh component follows from fluctuation-dissipation relations, 
$i\Sigma^K_{\rm eq}(E) = \tanh\left(\frac{E}{2T}\right)\Gamma_{\rm eq}(E)$.

For general perturbations from equilibrium $g^K=g^K_{\rm eq}+\phi$, the self energies are linearly perturbed:
\begin{equation}
\begin{aligned}
\Sigma^X[g^K(E)]-\Sigma^X_{\rm eq}(E)&=\int D^X(E')\delta g^K(t,E-E') \frac{dE'}{2\pi}\\ &\equiv \hat{D}^X\delta g^K (t,E)
\end{aligned}
\end{equation}
where $X$ stands for $R/A/K$. Using this, the linearized version of the KBE is 
\begin{equation}
\begin{aligned}
\partial_t\phi &+i\bigl[\Sigma^R_{\rm eq}\circ \phi-\phi\circ \Sigma^A_{\rm eq}\\ &+\bigl(\hat{D}^R\phi\bigr)\circ g^K_{\rm eq}-g^K_{\rm eq}\circ \bigl(\hat{D}^A\phi\bigr) -2\hat{D}^K \phi \bigr] =0.
\end{aligned}
\label{lin KBE phi}
\end{equation}

For a monochromatic perturbation as in Eq.\eqref{perturbation def}, i.e. $\phi(t,E)=\epsilon_\omega(t)\delta g^K(E)$ with $\epsilon_\omega(t)=\epsilon_0 e^{-i\omega t}$, all Moyal products in  Eq.~\eqref{lin KBE phi} are of the form:
\begin{equation}
\begin{aligned}
A(E)\circ (e^{-i\omega t}B(E)) = A\bigl(E+\frac{\omega}{2}\bigr)B(E), \\  
   (e^{-i\omega t}B(E))\circ A(E)= A\bigl(E-\frac{\omega}{2}\bigr)B(E).
\end{aligned}
\end{equation}
Henceforth we denote $A_\pm \equiv A_{\rm eq}\left(E\pm\frac{\omega}{2}\right)$ for various two-point functions. This gives the linearized KBE's kinetic operator at frequency $\omega$:

\begin{equation}
\begin{aligned}
\hat{K}_\omega &= -i\omega +\bigl(\Gamma_0-2i\hat{D}^K\bigr)+ i\bigl(\sigma^R_+-\sigma^A_-\bigr)\\ &\quad + i\bigl(g^K_-\hat{D}^R- g^K_+ \hat{D}^A\bigr).
\end{aligned}
\label{lin KBE pm}
\end{equation}
Since many of the constituent functions have poles at the Matsubara frequencies $-i\omega=2\pi(n+\tfrac{1}{2})T$, analytically continuing this result outside the principal strip $|{\rm Im}\omega|<2\pi T$ requires additional care. In Appendix~\ref{app: mer cont} we show that this analytical continuation of Eq.~\eqref{lin KBE pm} to arbitrary ${\rm Im}\omega$ requires either a coordinate change, or a subtle contour shift, and provide the generalized form of the KBE's kinetic operator extendable outside the principal strip.

In the $\Omega_0/T\to0$ limit, 
\begin{align}
    \Gamma_0-2i\hat{D}^K &\to -
    \pi g^2T\partial_E^2 \\ \nonumber  
    \hat{\mu} &\to  -\frac{\pi g^2}{2} \partial_E  \\ \nonumber 
    \sigma^{R/A}(E)&\to g^2\left[\partial_E \psi\left(\frac{1}{2}\mp i\frac{E}{2\pi T}\right)\right] 
\end{align}
and plugging into Eq.~\eqref{lin KBE pm} the kinetic operator becomes the continuous operator:
\begin{widetext}
\begin{equation}
\begin{aligned}
\hat{K}_\omega &= -i\omega - \pi g^2\bigl(  T\partial_E^2+\frac{i}{\pi}\bigl[\partial_E\bigl(\psi\bigl(\frac{1}{2}- i\frac{E+\omega/2}{2\pi T}\bigr)- \psi\bigl(\frac{1}{2}+ i\frac{E-\omega/2}{2\pi T}\bigr)\bigr)\bigr]\\ &\quad + \frac{1}{2}\bigl(g^K_-\mathcal{P}_+ + g^K_+\mathcal{P}_- \bigr)\partial_E \bigr)
\end{aligned}
\label{Kw end mat}
\end{equation}
\end{widetext}
 We have used $\pm i \hat{D}^{R/A}=\mathcal{P}_\pm \hat{\mu}$
where $\mathcal{P}_\pm =\frac{1\pm i\hat{\mathcal{H}}}{2}$ are causal/anticausal projections (i.e. the Fourier transforms of $\theta(\pm\tau)$), with $\hat{\mathcal{H}}$ the Hilbert-transform operator 
\begin{equation}
    \hat{\mathcal{H}}f(E) \equiv \frac{2}{E}*f(E)= \frac{1}{\pi}{\rm PV}\int dE' \frac{f(E')}{E-E'}.
\end{equation}

% Supplemental Material stays in this source for the combined arXiv version.
\clearpage
\onecolumngrid
\setcounter{page}{1}
\renewcommand{\thepage}{S\arabic{page}}
\setcounter{equation}{0}
\renewcommand{\theequation}{S\arabic{equation}}
\setcounter{figure}{0}
\renewcommand{\thefigure}{S\arabic{figure}}
\setcounter{table}{0}
\renewcommand{\thetable}{S\arabic{table}}
\setcounter{section}{0}
\setcounter{subsection}{0}
\setcounter{subsubsection}{0}
\setcounter{secnumdepth}{3}
\renewcommand{\thesubsection}{S\arabic{subsection}}
\renewcommand{\thesubsubsection}{\alph{subsubsection}}
\section*{Supplemental Material}

%\subsection{Details of numerical calculations}

\subsection{Analytic continuation of $\hat{K}_\omega$ outside the principal strip $|{\rm Im}\omega|<2\pi T$}
\label{app: mer cont}

Recall the linearized KBE for the deviation from equilibrium written as  $\phi(t,E)=g^K(t,E)-g^K_{\rm eq}(E)$:

\begin{equation}
    \partial_t\phi +i\left[ \Sigma^R_{\rm eq}\circ \phi-\phi\circ \Sigma^A_{\rm eq}+\left(\hat{D}^R\phi\right)\circ g^K_{\rm eq}-g^K_{\rm eq}\circ \left(\hat{D}^A\phi\right)-2\hat{D}^K \phi \right] =0.
\end{equation}

For a real-frequency solution $\phi(t,E)\sim e^{-i\omega t}\delta g^K(E)$ this can be simplified by shifting appropriate equilibrium functions by $E\to E\pm \omega/2$, as shown in Eq.~\eqref{lin KBE pm} of the End Matter. This method can be trivially extended  to complex $\omega$ with $-2\pi T<{\rm Im}\omega<2\pi T$ by shifting some intermediate integration contours. However, once $\omega$ exceeds this strip, $E\pm \omega/2$ pass through the lowest Matsubara frequencies $E=\pm i\pi T$ which are poles of the equilibrium functions $\Sigma^{R/A}_{\rm eq},g^K_{\rm eq}$ and the naive derivation of this equation fails. In this appendix we set out to understand this failure and how to bypass it.

Formally, the KBE is a particular instance of a ``retarded functional differential equation'' (RFDE) \cite{hale1993functional} . We will begin by a discussion of generic RFDEs and the (harmless) singularities encountered there, and then point at the additional complications which emerge due to the two-time structure of the KBE. 

To begin with, consider $\vec{\phi}(t)$ an element of some vector space $V$ (which could be  either finite or infinite dimensional). We define a general RFDE with kernel $\underline{h}(t):V\to V$ (analogous to the collision operator):
\begin{equation}
    \frac{d \vec{\phi}(t)}{dt}=\int_0^\infty \underline{h}(s)\vec{\phi}(t-s) ds.
\end{equation}

It is instructive to consider two different notions of solutions of such an RFDE:
\begin{itemize}
    \item \textbf{Eigenvectors:} We can consider solutions of the form $\vec{\phi}(t) = e^{\gamma t}\vec{\phi}_\gamma$ , extending to all times $t\in (-\infty,\infty)$. Plugging this form into the RFDE, we arrive at the characteristic equation
\begin{equation}
    \left(\gamma - \underline{h}(\gamma) \right) \vec{\phi}_\gamma = 0 \label{characteristic equation}
\end{equation}
where $\underline{h}(\gamma) = \int_0^\infty \underline{h}(t)e^{-\gamma t}dt$ is the Laplace transform. 
This integral definition is valid only in the half plane ${\rm Re}\gamma > {\rm Re}\gamma_0$, where $\gamma_0$ is the pole of $\underline{h}(\gamma)$ with maximal real part (e.g. for $h(t)=e^{-t}$, $\gamma_0=-1$). Given the analytic form of $\underline{h}(\gamma)$ the characteristic equation \textit{can} be analytically continued to the whole complex plane (excluding poles of $\underline{h}(\gamma)$). However, eigenvectors with $\gamma_*$ lying in the continued half plane cannot be regarded as genuine solutions of the RFDE, since the integral $\int_0^\infty \underline{h}(s)\vec{\phi}(t-s)ds\sim e^{\gamma_*t}\int_0^\infty e^{(\gamma_0-\gamma_*)s}ds$ is divergent.

\item \textbf{Fundamental solution:} An initial-value problem (IVP) is well defined given the fixed history $\vec{\phi}(t\leq0)$ as input. Consider the solution $\vec{\phi}(t)$ for a history $\vec{\phi}(t<0)=0,\vec{\phi}(t=0)=\vec{\phi}_0$. The RFDE becomes
\begin{equation}
    \frac{d \vec{\phi}(t)}{dt}=\int_0^t \underline{h}(s)\vec{\phi}(t-s) ds.
\end{equation}
Since all constituent functions are nonzero only for $t\geq 0$, this version of the RFDE is amenable to Laplace-transformation, and we arrive at the Laplace-transform of the solution:
\begin{equation}
\vec{\phi}(\gamma)=\left(\gamma - \underline{h}(\gamma) \right)^{-1} \vec{\phi}_0,
\end{equation}
from which the fundamental solution is defined as:
\begin{equation}
    \underline{\Phi}(\gamma) = \left(\gamma - \underline{h}(\gamma) \right)^{-1}. \label{fundamental solution}
\end{equation}
The fundamental solution serves as a Green's function, and solutions with arbitrary histories can be recovered from the fundamental solution using the variation-of-constants formula \cite{hale1993functional} (with a nonzero history at $t<0$ entering as an external forcing at times $t>0$). The long-time dynamics of any generic solution to the IVP are determined by poles of the fundamental solution.
\end{itemize}

Notice that poles of the fundamental solution Eq.~\eqref{fundamental solution} correspond exactly to solutions of the characteristic equation Eq.~\eqref{characteristic equation}, extended to the whole complex plane.
Thus, even if a solution of the characteristic equation in the analytically continued range ${\rm Re \gamma_*}<{\rm Re}\gamma_0$ does not correspond to a well defined solution $e^{\gamma t}\vec{\phi}_\gamma$ for all $t\in (-\infty,\infty)$, it still determines a physically meaningful resonance of IVPs. Another way to look at this is that for a solution of the characteristic equation with ${\rm Re}\gamma_*>{\rm Re}\gamma_0$, we can prepare a history $\vec{\phi}(t<0)=e^{\gamma_* t}\vec{\phi}_{\gamma_*}$ which extends to $t>0$ by solution of the IVP. On the other hand, for eigenvalues ${\rm Re}\gamma_*<{\rm Re}\gamma_0$ there is no way to prepare a history which will result in a single eigenvector solution for all $t>0$.

This motivates us to construct the analogue of the characteristic equation for the KBE and extend it to arbitrary $\gamma$, without need for concern for the fact that we extend beyond a pole of the kernel (i.e. the collision operator). However, as we will now show, the singularities in the KBE's kinetic operator are not purely of the aforementioned type, and thus further care is required.

\begin{figure}[t]
\centering
\includegraphics[width=0.7\textwidth]{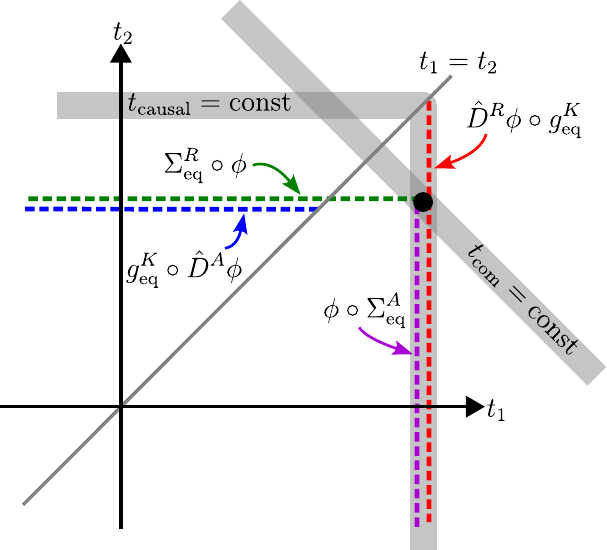} 
\caption{Sketch of the convolution trajectories of the non-local terms in the linearized KBE. The dashed lines indicate the domain of $\phi(t_1,t_2)$ probed by the convolutions which contribute to the collision integral at the point marked by the black dot.}
\label{fig: convolutions}
\end{figure}

In the context of the KBE,  constructing the characteristic equation rather than the fundamental solution has the advantage of allowing us to work with the center-of-mass (com) time coordinate. To see this, consider the two different candidates for a ``central'' time coordinate:
\begin{align}
    t_{\rm com} &= \frac{t_1+t_2}{2} \\ \nonumber 
    t_{\rm causal} &= \max(t_1,t_2).
\end{align}
($t_{\rm com}$ is what we have been calling $t$ up to this point.)
In both cases the relative time coordinate is taken to be $\tau=t_1-t_2$.  While $t_{\rm com}$ is a convenient time coordinate for most manipulations we have performed on the KBE, it is somewhat problematic as a time coordinate: the KBE is \textit{not causal} with respect to $t_{\rm com}$. This can be seen in Fig.~\ref{fig: convolutions}, where we have plotted the convolution ranges of the different terms contributing to the linearized collision operator at a given point $(t_1,t_2)$, compared to the surfaces of constant $t_{\rm com},t_{\rm causal}$ to which this point belongs. A physical IVP is thus only well defined using $t_{\rm causal}$, for which the collision integral only involves points with $t_{\rm causal}$ lower than that of the given point. 

On the other hand, when constructing the characteristic equation we use solutions which extend throughout the whole $(t_1,t_2)$ plane. These types of solutions can be equivalently written using either $t_{\rm com}$ or $t_{\rm causal}$:
\begin{align}
    \phi(t_1,t_2) = e^{\gamma t_{\rm com}}\phi_{\gamma}^{(\rm com)}(\tau)= e^{\gamma t_{\rm causal}}\phi_{\gamma}^{(\rm causal)}(\tau)
\end{align}
and since we are not solving an IVP, we have no reason to enforce any notion of causality in our methods (beyond the existing causal structure in the Keldysh time contour). Using $t_{\rm causal}=t_{\rm com}+|\tau|/2$, we can relate the two $\phi_\gamma(\tau)$ by a basis change:
\begin{equation}
    \phi_{\gamma}^{(\rm com)}(\tau)= e^{\gamma |\tau|/2}\phi_{\gamma}^{(\rm causal)}(\tau)
    \label{phi relation}
\end{equation}

We begin by focusing on the singularity of the characteristic equation at ${\rm Re}\gamma=2\pi T$. Since all temporal correlations encoded in the equilibrium functions $g^K_{\rm eq},\Sigma^{R/A}_{\rm eq}$ decay with time separation,  the analogue of the kernel $\underline{h}(t)$ decays with $t_{\rm com}$, and the first pole of its Laplace-transform should be regarded as at $\gamma_0 = -2\pi T$. Thus, the singularity at ${\rm Re}\gamma =2\pi T$ is different than that mentioned earlier, at which eigenvector solutions cease to be well defined solutions.

How can we understand this singularity? It is instructive to examine the effect of the term which reaches into ``future'' values of $t_{\rm com}$, e.g. for $\tau>0$ the red term in Figure~\ref{fig: convolutions} (for $\tau<0$ it exchanges roles with the blue term $g^K_{\rm eq}\circ D^A\phi$). For ${\rm Re}\gamma >0$, solutions grow with time, and thus this  term should become increasingly important. Plugging in an eigenvector solution, we find that for large $\tau$ this term is schematically of the form 
\begin{align}
    \left(D^R \phi \circ g^K_{\rm eq}\right) (t_1,t_2)& = \int D^R(t_1-t_3)\phi(t_1,t_3)g^K_{\rm eq}(t_3-t_2)dt_3 \\ \nonumber  &= \int_0^\infty D^R(s) \phi\left(t_{\rm com}+\frac{\tau-s}{2},s\right) g_{\rm eq}^K(\tau -s)ds\\ \nonumber 
    & \propto  e^{\gamma t_{\rm com}} e^{(\gamma/2 - \pi T )\tau} \int_0^\tau e^{(-\gamma/2 \pm i\Omega_0 )s} e^{\pi T s} \phi_{\gamma}^{(\rm com)}(s) ds + \cdots
\end{align}
using for large arguments $g^K_{\rm eq}(\tau )\propto e^{-\pi T \tau }$. Peeling off the common prefactor $e^{\gamma t_{\rm com}}$, we find that although the integral itself could be completely well behaved and converge to some finite value with $\tau \to \infty$, the prefactor $e^{(\gamma/2-\pi T)\tau}$ becomes dangerous when ${\rm Re}\gamma >2\pi T$; this term can no longer be regarded as an operator closed on $L^2$ (normalizable) functions. The singularity in the kinetic operator arising from this term signals that  analytically  continuing the Fourier transform (FT) of this term with respect to $\tau$ on the real axis will fail to represent it reliably\footnote{We note that, to the best of our knowledge, there is no fundamental constraint preventing the eigenfunctions from having tails that grow exponentially with $|\tau|$. Furthermore, it is not difficult to construct examples where an RFDE has eigenfunctions with exponentially growing tails, but the solution of any IVP with $L^2$ initial conditions remains $L^2$ at any finite time. A simple example similar in spirit to the KBE is $df(t,\tau)/dt = -\int_0^\infty e^{-(\tau+s)}f(t-\tau,s)ds$}.

This is equivalent to saying that the retarded function $\theta(\tau)e^{-a\tau}$ can only be represented by a FT on the real axis as long as ${\rm Re}a>0$. If we forcibly continue its FT $(a-iE)^{-1}$  to the half plane ${\rm Re}a<0$ and invert on the real axis, we end up with the advanced function  $-\theta(-\tau)e^{+a\tau}$ instead. 
How can we deal with this singularity? We can either shift the transformation contour from the real axis, so as to be able to accommodate functions with exponentially growing tails, or equivalently, using Eq.~\ref{phi relation} change to the $t_{\rm causal}$ 
basis which keeps this term well behaved:
\begin{align}
     e^{(\boldsymbol{\gamma/2} - \pi T )\tau} \int_0^\tau e^{(-\boldsymbol{\gamma/2}  \pm i\Omega_0 )s} e^{\pi T s} \phi_{\gamma}^{(\rm com)}(s) ds \leftrightarrow e^{ - \pi T \tau} \int_0^\tau e^{ \pm i\Omega_0 s} e^{\pi T s} \phi_{\gamma}^{(\rm causal)}(s) ds
\end{align}
Similar behavior can be found for $\tau<0$, replacing $\tau \to |\tau|$. Thus, changing to the $t_{\rm causal}$ basis ``untangles'' the problematic $\gamma$ exponents from these terms, and they cease to be singular (this makes sense since these terms lie on a constant $t_{\rm causal}$ slice, as shown in Figure~\ref{fig: convolutions}).
Deriving the form of the operator using the $t_{\rm causal}$ basis is significantly messier. To do it, we apply the FT of the transformation $e^{-\gamma|\tau|/2}$ in the unproblematic region $2\pi T>\rm{Re}\gamma>0$ and then after having removed the singularity at ${\rm Re}\gamma = 2\pi T$ extend to larger ${\rm Re}\gamma$. For $\mathcal{F}$ the FT operator, we define
\begin{equation}
    U_\gamma = \mathcal{F} e^{-\gamma|\tau|/2}\mathcal{F}^{-1}
\end{equation}
and we wish to find
\begin{equation}
    \hat{K}_\gamma^{(\rm causal)} = U_\gamma\hat{K}_\gamma^{(\rm com)}U_\gamma^{-1}
\end{equation}
with $\hat{K}_\gamma^{(\rm com)}$ as defined in Eq.~\eqref{lin KBE pm} of the End Matter (with $\omega \to i\gamma$).
Considering its action only on functions $f(E)$ which do not have poles in the strip $|{\rm Im }E|<\pi T$, we can write
\begin{align}
    U_\gamma f(E)= f_+(E+i\gamma/2) +f_-(E-i\gamma/2)  
\end{align}
where $f_\pm (E)$ are the FT of $\theta(\pm\tau)f(\tau)$, with $f_+$ analytic in the UHP, and $f_-$ analytic in the LHP. We will make use of the projectors:
\begin{align}
    \mathcal{P}_\pm &= \mathcal{F} \theta(\pm \tau)\mathcal{F}^{-1} \\ 
    \mathcal{P}_\pm f(E) &= f_\pm (E) \nonumber
\end{align}
which remove all poles in the upper/lower half planes,  and shifts 
\begin{equation}
    T_{\pm i\gamma/2} f(E)\equiv f(E\pm i\gamma/2)
\end{equation}
(which shift all poles downwards/upwards). As explained above, we are assuming ${\rm Re}\gamma<2\pi T$, and further considering only action on functions with no poles in the strip $|{\rm Im }E|<\pi T$, on which  the actions of $\mathcal{P}_{\pm},T_{\pm i\gamma/2}$ commute. We can organize the two components in vector form:
\begin{equation}
    f(E)\to \begin{pmatrix}
        f_+(E) \\    f_-(E)
    \end{pmatrix}
\end{equation}
and then represent
\begin{align}
    U_\gamma &= \begin{pmatrix}
        T_{+i\gamma/2} & 0 \\   0 & T_{-i\gamma/2}
    \end{pmatrix}, \quad
    U^{-1}_\gamma = \begin{pmatrix}
        T_{-i\gamma/2} & 0 \\   0 & T_{+i\gamma/2}
    \end{pmatrix} .
\end{align}
We now derive conjugation of the various constituent functions and operators of Eq.~\eqref{lin KBE pm} by $U_\gamma$. To begin with, all phonon propagators are diagonal in $\tau$ and thus, as operators, commute with $U_\gamma$. To derive the transformation of the other terms, we begin by considering the transformation of an operator that attaches a simple pole: $d_a f(E) \equiv \frac{1}{E-a}f(E)$. For $a\in$UHP, $d_af_-(E)$ remains analytic in the LHP, i.e. $\mathcal{P}_- d_a f_- = d_a f_-$. Its action on the other component can be decomposed:
\begin{align}
    \frac{1}{E-a}f_+(E) = \frac{f_+(E)-f_+(a)}{E-a} + \frac{f_+(a)}{E-a}
\end{align}
where $f_+(a)$ is well defined since $f_+$ is analytic in the UHP. Thus the first term is analytic in the UHP, and the 2nd term is analytic in the LHP. Defining the operator $\delta_a f_+(E)=f_+(a)$, we can summarize:
\begin{equation}
    d_{a\in {\rm UHP}} = \frac{1}{E-a}\begin{pmatrix}
      1-\delta_a & 0 \\   \delta_a & 1  
    \end{pmatrix}.
\end{equation}
Similarly, for $a\in$LHP we can decompose:
\begin{equation}
    d_{a\in {\rm LHP}} = \frac{1}{E-a}\begin{pmatrix}
      1 & \delta_a \\   0 & 1-\delta_a  
    \end{pmatrix},
\end{equation}
where, this time, $\delta_a$ only acts on the $f_-$ component. We can represent the action of $\delta_{a}$ on functions of real $E$ as
\begin{align}
    \delta_{a\in LHP}f_- &=  -\frac{1}{2\pi i} \int \frac{dE'}{E'-a} f_-(E') = -\frac{1}{2\pi i} \int \frac{dE'}{E'-a} f(E') \\ \nonumber  
    \delta_{a\in UHP}f_+ &=  \frac{1}{2\pi i} \int \frac{dE'}{E'-a} f_+(E') = \frac{1}{2\pi i} \int \frac{dE'}{E'-a} f(E') 
\end{align}
where in the last step we can remove the subscripts $\pm$ from $f$ inside the integrals since the integrals of the complementary components have all poles in a single half plane, and therefore vanish.

Then, conjugation by $U_\gamma$ gives (assuming that $|{\rm Im}a|>\pi T$):

\begin{align}
     U_\gamma d_{a\in {\rm UHP}} U_\gamma^{-1} &= \begin{pmatrix}
      T_{+i\gamma/2}\frac{1-\delta_a}{E-a}T_{-i\gamma/2} & 0 \\   T_{-i\gamma/2}\frac{\delta_a}{E-a}T_{-i\gamma/2} & T_{-i\gamma/2}\frac{1}{E-a} T_{+i\gamma/2} 
    \end{pmatrix} = \begin{pmatrix}
      \frac{1-\delta_{a-i\gamma/2}}{E+i\gamma/2-a} & 0 \\   \frac{\delta_{a-i\gamma/2} }{E-i\gamma/2-a} & \frac{1}{E-i\gamma/2-a} 
    \end{pmatrix} \\ \nonumber  
    U_\gamma d_{a\in {\rm LHP}} U_\gamma^{-1} &= \begin{pmatrix}
      \frac{1}{E+i\gamma/2-a} & \frac{\delta_{a+i\gamma/2} }{E+i\gamma/2-a} \\ \nonumber  0 & \frac{1-\delta_{a+i\gamma/2}}{E-i\gamma/2-a}
    \end{pmatrix}.
\end{align} 

From this we can derive the conjugation of the digamma function, using its representation as:
\begin{align}
    \psi\left(\frac{1}{2}+\frac{iE}{2\pi T}\right) -\psi\left(\frac{1}{2}\right)&= N\sum_{\omega_n>0} \frac{1}{i\omega_n -E} - \frac{1}{i\omega_n} = -N\sum_{\omega_n>0}d_{i\omega_n }  - \frac{1}{i\omega_n}
    \\ \nonumber  \psi\left(\frac{1}{2}-\frac{iE}{2\pi T}\right) -\psi\left(\frac{1}{2}\right)&= N\sum_{\omega_n>0} \frac{1}{i\omega_n +E}  - \frac{1}{i\omega_n} = N\sum_{\omega_n>0}d_{-i\omega_n } - \frac{1}{i\omega_n}
\end{align}
with the sum running over positive Matsubara frequencies $\omega_n=(2n+1)\pi T$, and $N=-2\pi i T$. The combinations we need are:
\begin{align}
    U_\gamma \psi\left(\frac{1}{2}+\frac{i(E - i\gamma/2)}{2\pi T}\right)U_\gamma^{-1} -\psi\left(\frac{1}{2}\right)&= -N \sum_{\omega_n>0} U_\gamma d_{i\omega_n+ i\gamma/2}(E)U_\gamma^{-1} - \frac{1}{i\omega_n}\nonumber\\ &= N \sum_{\omega_n>0}\begin{pmatrix}
        \frac{1-\delta_{i\omega_n}}{i\omega_n-E} & 0 \\   \frac{\delta_{i\omega_n}}{ i\omega_n +i\gamma -E} & \frac{1}{ i\omega_n +i\gamma -E } 
    \end{pmatrix} - \frac{1}{i\omega_n} \\ \nonumber 
    &\to  N \sum_{\omega_n>0}\left[\left(\frac{1-\delta_{i\omega_n}}{i\omega_n-E}+\frac{\delta_{i\omega_n}}{ i\omega_n +i\gamma -E}\right)\mathcal{P}_+ + \frac{1}{ i\omega_n +i\gamma -E }\mathcal{P}_-  - \frac{1}{i\omega_n}\right] \\ \nonumber 
    U_\gamma \psi\left(\frac{1}{2}-\frac{i(E + i\gamma/2)}{2\pi T}\right)U_\gamma^{-1}-\psi\left(\frac{1}{2}\right) &= N \sum_{\omega_n>0} U_\gamma d_{-i\omega_n- i\gamma/2}(E)U_\gamma^{-1} - \frac{1}{i\omega_n}\nonumber\\ \nonumber &= N \sum_{\omega_n>0}\begin{pmatrix}
         \frac{1}{ i\omega_n +i\gamma +E } & \frac{\delta_{-i\omega_n}}{ i\omega_n +i\gamma +E}\\ \nonumber  0 & \frac{1-\delta_{-i\omega_n}}{i\omega_n+E}
    \end{pmatrix}  - \frac{1}{i\omega_n}\\ \nonumber 
    &\to  N\sum_{\omega_n>0} \left[\frac{1}{ i\omega_n +i\gamma +E }\mathcal{P}_+ + \left(\frac{1-\delta_{-i\omega_n}}{i\omega_n+E}+\frac{\delta_{-i\omega_n}}{ i\omega_n +i\gamma +E}\right)\mathcal{P}_- - \frac{1}{i\omega_n}\right].
\end{align}
Working in the operator convention $A*f=\int A(E,E')f(E')\frac{dE'}{2\pi}$ we can replace $\delta_{a}\mathcal{P}_\pm \to \frac{i}{a\mp E'}$ and, using partial fractions when necessary, recombine the sums into new digamma functions ($\Delta E = E-E'$):
\begin{align}
    U_\gamma& \psi\left(\frac{1}{2}+\frac{i(E - i\gamma/2)}{2\pi T}\right)U_\gamma^{-1} -\psi\left(\frac{1}{2}\right)\\ \nonumber&= N\sum_{\omega_n>0} \frac{1}{i\omega_n-E}\left(\mathcal{P}_+ - \frac{i}{\Delta E}\right)+ \frac{1}{i\omega_n-(E-i\gamma)}\left(\mathcal{P}_- + \frac{i}{\Delta E-i\gamma}\right) + \frac{1}{i\omega_n-E'} \left( \frac{i}{\Delta E}-\frac{i}{\Delta E -i\gamma}\right)  - \frac{1}{i\omega_n}\\ \nonumber  &= \psi\left(\frac{1}{2}+\frac{iE}{2\pi T}\right)\left(\mathcal{P}_+ - \frac{i}{\Delta E}\right)+  \psi\left(\frac{1}{2} + \frac{\gamma}{2\pi T} +\frac{iE}{2\pi T}\right)\left(\mathcal{P}_- + \frac{i}{\Delta E-i\gamma}\right) + \psi\left(\frac{1}{2}+\frac{iE'}{2\pi T}\right) \left( \frac{i}{\Delta E}-\frac{i}{\Delta E - i\gamma}\right) \nonumber\\ &=\psi\left(\frac{1}{2} + \frac{\gamma}{2\pi T} +\frac{iE}{2\pi T}\right)\left(\mathcal{P}_- + \frac{i}{\Delta E-i\gamma}\right) + \psi\left(\frac{1}{2}+\frac{iE'}{2\pi T}\right) \left( \mathcal{P}_+-\frac{i}{\Delta E - i\gamma}\right) \nonumber \\ 
    U_\gamma &\psi\left(\frac{1}{2}-\frac{i(E + i\gamma/2)}{2\pi T}\right)U_\gamma^{-1}  \nonumber-\psi\left(\frac{1}{2}\right)\\&=\psi\left(\frac{1}{2} + \frac{\gamma}{2\pi T} -\frac{iE}{2\pi T}\right)\left(\mathcal{P}_+ - \frac{i}{\Delta E+i\gamma}\right) + \psi\left(\frac{1}{2}-\frac{iE'}{2\pi T}\right) \left( \mathcal{P}_-+\frac{i}{\Delta E + i\gamma}\right)  \nonumber
\end{align}
(the last row can be either re-derived in the same way, or can be obtained by assuming real $\gamma$ and taking the complex conjugate of the previous result). Note that in this convention, the projectors are $\mathcal{P}_\pm = \pi \delta(\Delta E) \pm \frac{i}{\Delta E}$.

The inelastic components of the self energies can be expressed as 
\begin{equation}
    \sigma^{R/A} = \frac{g^2}{2\Omega_0 } \left(T_{+\Omega_0} \psi\left(\frac{1}{2} \mp \frac{iE}{2\pi T}\right) T_{-\Omega_0}-T_{-\Omega_0} \psi\left(\frac{1}{2} \mp \frac{iE}{2\pi T}\right) T_{+\Omega_0}\right)
\end{equation}
and since $T_{\pm \Omega_0}$ commute with $U_\gamma$ we can interchange their order. This takes care of transforming the diagonal term in Eq.~\eqref{lin KBE pm}:
\begin{align}
U_\gamma &i\left(\sigma^R(E+i
    \gamma/2) - \sigma^A(E-i\gamma/2)\right) U_\gamma^{-1} =  \frac{\pi g^2}{2\Omega_0}\left(\Psi_\gamma(E-\Omega_0,E'-\Omega_0)-\Psi_\gamma(E+\Omega_0,E'+\Omega_0)\right) \\
    \Psi_\gamma(E,E') &= \frac{i}{\pi}\left[\psi\left(\frac{1}{2} + \frac{\gamma}{2\pi T} -\frac{iE}{2\pi T}\right)\left(\mathcal{P}_+ - \frac{i}{\Delta E+i\gamma}\right) + \psi\left(\frac{1}{2}-\frac{iE'}{2\pi T}\right) \left( \mathcal{P}_-+\frac{i}{\Delta E + i\gamma}\right) \right.  \nonumber\\ &\left. -\psi\left(\frac{1}{2} + \frac{\gamma}{2\pi T} +\frac{iE}{2\pi T}\right)\left(\mathcal{P}_- + \frac{i}{\Delta E-i\gamma}\right) - \psi\left(\frac{1}{2}+\frac{iE'}{2\pi T}\right) \left( \mathcal{P}_+-\frac{i}{\Delta E - i\gamma}\right)\right] \nonumber
\end{align}
The final set of terms we must transform is 
\begin{equation}
    i\left[g_{\rm eq}^K(E-i\gamma/2) \hat{D}^R - g_{\rm eq}^K(E+i\gamma/2) \hat{D}^A\right] =  \left[g^K_{\rm eq}(E-i\gamma/2) \mathcal{P}_+ + g^K_{\rm eq}(E+i\gamma/2) \mathcal{P}_-\right] \hat{\mu}
\end{equation}
with the phonon spectral operator
\begin{equation}
    \hat{\mu} = \frac{\pi g^2}{4\Omega_0}\left(T_{-\Omega_0}-T_{\Omega_0}\right).
\end{equation}
$\hat{\mu}$ commutes with $U_\gamma$, so it suffices to calculate the conjugation of $g^K_{\rm eq}(E\mp i\gamma/2)\mathcal{P}_{\pm}$. We can write 
\begin{equation}
    g^K_{\rm eq}(E) = 4T\sum_{\omega_n} \frac{1}{E-i\omega_n } = 4T\sum_{\omega_n>0} d_{i\omega_n}(E) + d_{-i\omega_n}(E)
\end{equation}
and play a similar game, inserting the intermediate projectors as necessary:
\begin{align}
    U_\gamma g^K_{\rm eq}(E-i\gamma/2) \mathcal{P}_+ U^{-1}_\gamma &=  4TU_\gamma \sum_{\omega_n>0} \left(d_{i\omega_n+i\gamma/2}(E) + d_{-i\omega_n+i\gamma/2}(E) \right)  U^{-1}_\gamma \mathcal{P}_+  \\ \nonumber  
    & = 4T\sum_{\omega_n>0} \left[ \begin{pmatrix}
      \frac{1-\delta_{i\omega_n}}{E-i\omega_n} & 0 \\ \nonumber  \frac{\delta_{i\omega_n} }{E-i\gamma-i\omega_n} & 0  
    \end{pmatrix} + \begin{pmatrix}
      \frac{1}{E+i\gamma+i\omega_n} & 0\\ \nonumber  0 & 0
    \end{pmatrix}   \right] \\ \nonumber 
    &\to 4T\sum_{\omega_n>0}  \left[ \left(\frac{1}{E-i\omega_n} + \frac{1}{E+i\gamma+i\omega_n}\right)\mathcal{P}_+ + \left(\frac{1}{E-i\gamma-i\omega_n}-\frac{1}{E-i\omega_n} \right)\delta_{i\omega_n}\mathcal{P}_+\right]\\ \nonumber 
     U_\gamma g^K_{\rm eq}(E+i\gamma/2) \mathcal{P}_- U^{-1}_\gamma &= 4T\sum_{\omega_n>0}  \left[ \left(\frac{1}{E+i\omega_n} + \frac{1}{E-i\gamma-i\omega_n}\right)\mathcal{P}_- + \left(\frac{1}{E+i\gamma+i\omega_n}-\frac{1}{E+i\omega_n} \right)\delta_{-i\omega_n}\mathcal{P}_-\right]
\end{align}
Combining both:
\begin{align}
    U_\gamma\big[g^K_{\rm eq}(E-i\gamma/2) \mathcal{P}_+ +& g^K_{\rm eq}(E+i\gamma/2) \mathcal{P}_-\big]U_\gamma^{-1} \\ \nonumber &= 4T \sum_{\omega_n>0}  
    \left[ \frac{1}{i\omega_n-E}\left(-\mathcal{P}_+ + \frac{i}{\Delta E}\right) + \frac{1}{i\omega_n+ E}\left(\mathcal{P}_- + \frac{i}{\Delta E} \right) \right. \\ \nonumber  
    &  + \frac{1}{i\omega_n+i\gamma-E}\left(-\mathcal{P}_- - \frac{i}{\Delta E-i\gamma}\right) + \frac{1}{i\omega_n+i\gamma+ E}\left(\mathcal{P}_+ - \frac{i}{\Delta E+i\gamma} \right)  \\ \nonumber  
    &+\left.   \frac{1}{i\omega_n-E'}\left(\frac{i}{\Delta E-i\gamma}-\frac{i}{\Delta E}\right) + \frac{1}{i\omega_n+ E'}\left(\frac{i}{\Delta E+i\gamma}-\frac{i}{\Delta E} \right) \right]
    \\ \nonumber &=\frac{2i}{\pi}  
    \left[ \psi\left(\frac{1}{2}+\frac{iE}{2\pi T}\right)\left(-\mathcal{P}_+ + \frac{i}{\Delta E}\right) + \psi\left(\frac{1}{2}-\frac{iE}{2\pi T}\right)\left(\mathcal{P}_- + \frac{i}{\Delta E} \right) \right. \\ \nonumber  
    &  + \psi\left(\frac{1}{2}+\frac{\gamma}{2\pi T}+\frac{iE}{2\pi T}\right)\left(-\mathcal{P}_- - \frac{i}{\Delta E-i\gamma}\right) +  \psi\left(\frac{1}{2}+\frac{\gamma}{2\pi T}-\frac{iE}{2\pi T}\right)\left(\mathcal{P}_+ -\frac{i}{\Delta E+i\gamma} \right)  \\ \nonumber  
    &+\left.   \psi\left(\frac{1}{2}+\frac{iE'}{2\pi T}\right)\left(\frac{i}{\Delta E-i\gamma}-\frac{i}{\Delta E}\right) + \psi\left(\frac{1}{2}-\frac{iE'}{2\pi T}\right)\left(\frac{i}{\Delta E+i\gamma}-\frac{i}{\Delta E} \right)  \right] 
     \\ \nonumber 
    &= \frac{2i}{\pi}  
    \left[-\psi\left(\frac{1}{2}+\frac{\gamma}{2\pi T}+\frac{iE}{2\pi T}\right)\left(\mathcal{P}_- + \frac{i}{\Delta E-i\gamma}\right) +  \psi\left(\frac{1}{2}+\frac{\gamma}{2\pi T}-\frac{iE}{2\pi T}\right)\left(\mathcal{P}_+ - \frac{i}{\Delta E+i\gamma} \right)\right.  \\ \nonumber  
    &-\left.   \psi\left(\frac{1}{2}+\frac{iE'}{2\pi T}\right)\left(\mathcal{P}_+-\frac{i}{\Delta E-i\gamma} \right) + \psi\left(\frac{1}{2}-\frac{iE'}{2\pi T}\right)\left(\mathcal{P}_-+ \frac{i}{\Delta E+i\gamma} \right) \right] \nonumber \\ 
    & = 2\Psi_\gamma(E,E') \nonumber
\end{align}
and thus
\begin{equation}
   U_\gamma \left[g^K_{\rm eq}(E-i\gamma/2) \mathcal{P}_+ + g^K_{\rm eq}(E+i\gamma/2) \mathcal{P}_-\right] \hat{\mu} U_\gamma^{-1}=  \frac{\pi g^2}{2\Omega_0}\left(\Psi_\gamma(E,E'+\Omega_0)-\Psi_\gamma(E,E'-\Omega_0)\right).
\end{equation}

For completeness, we present the full operator:
\begin{align}
 \label{lin KBE causal}    \hat{K}^{\rm causal}_\gamma(E,E')
&=  2\pi \gamma\delta(\Delta E)
+\frac{\pi g^2}{2\Omega_0}\Bigg[
2\pi\coth\left(\frac{\Omega_0}{2T}\right)
\bigg(2\delta(\Delta E)-\delta(\Delta E-\Omega_0)-\delta(\Delta E+\Omega_0)\bigg)
\\
&\qquad
+ \Psi_\gamma(E-\Omega_0,E'-\Omega_0)
- \Psi_\gamma(E+\Omega_0,E'+\Omega_0)
+ \Psi_\gamma(E,E'+\Omega_0)
- \Psi_\gamma(E,E'-\Omega_0)
\Bigg] . \nonumber
\end{align}

\begin{figure}[t]
\centering
\includegraphics[width=0.7\textwidth]{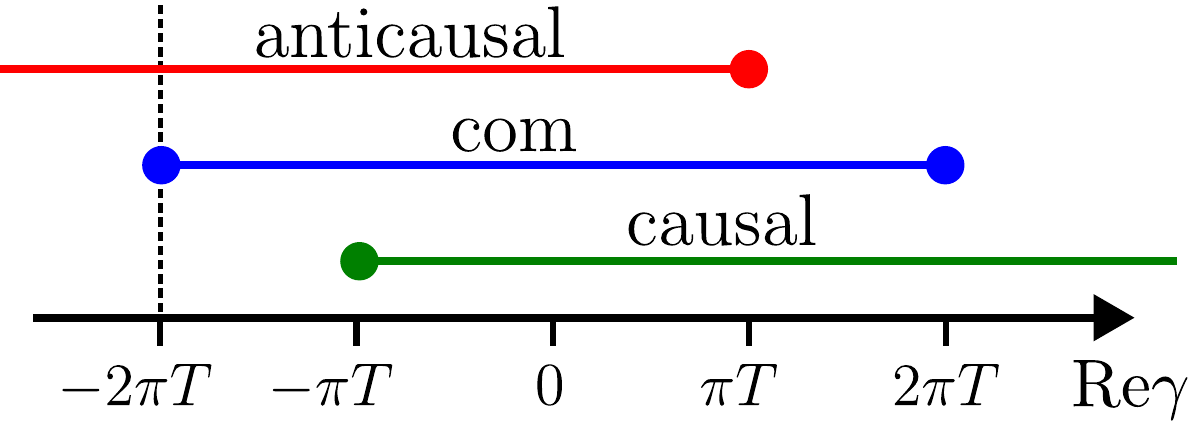} 
\caption{Comparison of the range of applicability of different choices of the central time coordinate, when searching for solutions of the characteristic equation using the Wigner transformed basis, i.e. solutions of the form $e^{\gamma t}\phi_\gamma(E)$. The Wigner transformed operator in the com basis is given in Eq.~\eqref{lin KBE pm} of the end matter, in the causal basis in Eq.~\eqref{lin KBE causal}, and the anticausal operator is not derived in this work. The dashed line refers to the ``harmless'' singularity at which the integral representation of eigenvectors ceases to hold, yet the characteristic equation remains valid.}
\label{fig: time coords}
\end{figure}

Recall that $\psi(z)$ has poles at $z=0,-1,-2,\cdots$. Thus, all above derived constituents are completely non-singular for ${\rm Re}\gamma >-\pi T$. Similarly to the singularity of the com operator at ${\rm Re}\gamma =2\pi T$ (which led us here in the first place), the singularity of the causal operator at ${\rm Re}\gamma =-\pi T$ also occurs since it starts to attach exponentially growing tails to $\phi(\tau)$. This can be seen explicitly by examining the terms extending backwards in $t_{\rm causal}$, and was expected since for negative ${\rm Re} \gamma$ the map $e^{-\gamma|\tau|/2}$ becomes problematic. 

This makes clear the fact that different choices of the central time coordinate are appropriate for solving the characteristic equation at different values of $\gamma$, as we illustrate in Figure~\ref{fig: time coords}. We note that both $t_{\rm com},t_{\rm causal}$ are inappropriate for eigenvalues with ${\rm Re}\gamma <-2\pi T$. In this work we were only interested in the slowest relaxing modes, which happened to never cross this limit, and thus do not delve into details regarding this region. We briefly note that this region can instead be addressed using the ``anticausal'' coordinate $t_{\rm anticausal}=\min\left(t_1,t_2\right)$, related by $\phi_\gamma^{(\rm anticausal)}(\tau)=e^{\gamma|\tau|/2}\phi_{\gamma}^{(\rm com)}(\tau)$. This coordinate should behave like a mirror of the causal coordinate, its Wigner transformed characteristic equation valid in the half plane ${\rm Re}\gamma<\pi T$. We expect that a derivation of the coordinate transformation/contour shift should be close in spirit to the one presented above. 

Finally, we note that while the operators $\hat{K}^{(\rm com)}_\gamma,\hat{K}^{(\rm causal)}_\gamma$ have exactly the same spectrum for $-\pi T< {\rm Re \gamma}<2\pi T$, any truncation and discretization of the energy axis $E\in \mathbb{R}\to E\in [-E_{\rm max},-E_{\max}+dE,\cdots,E_{\rm max}]$ will lead to errors.
Diagonalization of the naively discretized Eq.~\eqref{lin KBE causal} results in large errors even for very small, nonzero $\gamma$, and consequently the spectra of the two operators deviate from each other. This is due to the fact that the operator $\hat{K}^{(\rm com)}_\gamma$ is extremely ill-conditioned, e.g. at $\gamma=0$ the left eigenvectors are of the form $\sim \cosh(E/2T)$ at large $E$ (as shown in Appendix.~\ref{app: relaxation}). Since the basis transformation $U_\gamma$ has a kink at $\tau=0$, it has power law decaying tails in $E$ which do not bode well for the exponentially growing left eigenvectors.
While this would be avoided by using a smooth transformation such as $e^{\gamma\sqrt{\tau^2+a^2}/2}$ instead of $e^{\gamma|\tau|/2}$ (for an appropriately chosen $a$), it is much more complicated to apply it on the analytical form of the operator required for continuation outside the problematic $\gamma$ range.

\begin{figure}[t]
\centering
\includegraphics[width=0.7\textwidth]{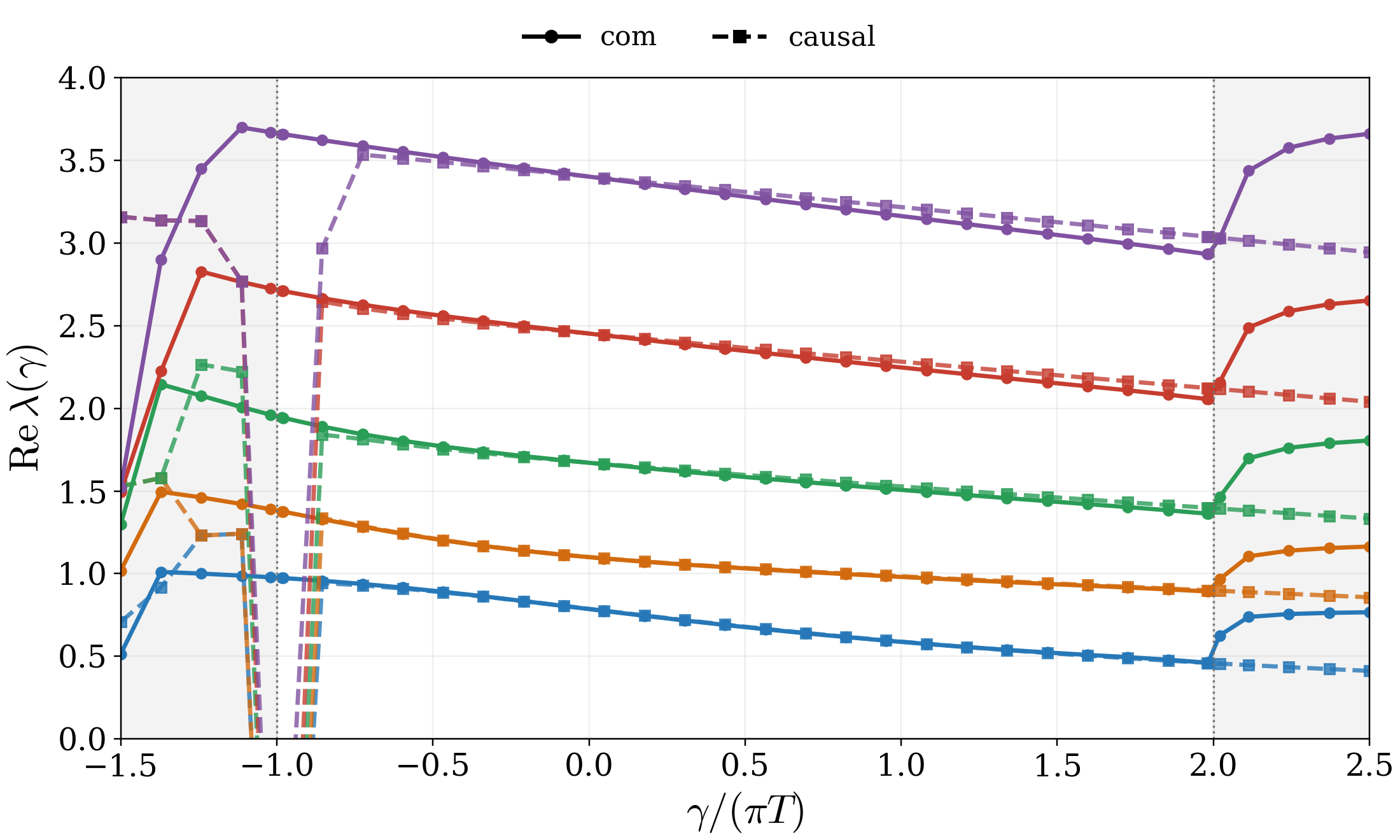} 
\caption{Comparison of the 5 lowest eigenvalues $\lambda(\gamma)$ of $\hat{C}_\gamma$ in the $t_{\rm com}$ and $t_{\rm causal}$ representations. For the causal version of the operator we use $E_1/T=30$ and then truncate down to $E_2/T=15$, with $a=4/T$ in the smoothing $V_\gamma$. As can be seen, the spectrum agrees exceptionally well in most of the overlapping region $-\pi T < {\rm Re}\gamma < 2\pi T$, with the causal version diverging near its limit at $\gamma = -\pi T$ and the com version jumping sharply at $\gamma =2\pi T$.}
\label{fig: caus com eigs}
\end{figure}

A simple solution which we found to this problem is to discretize Eq.~\eqref{lin KBE causal} on a larger energy axis (with $E_{{\rm max}}=E_1$), and then apply a similarity transformation by conjugating with the operator $V_\gamma = \mathcal{F}\exp\left(\gamma(|\tau|-\sqrt{\tau^2+a^2})/2\right)\mathcal{F}^{-1}$ represented on the discretized energy grid. Since this is a similarity transformation, it does not alter the spectrum of $\hat{K}^{\rm (causal)}_\gamma$, but then truncation of this operator to a smaller energy axis with $E_{\rm max}=E_2<E_1$ leads to a well behaved spectrum which agrees with the spectrum of $\hat{K}_\gamma^{(\rm com)}$ in the required $\gamma$ range, and can then be used to extend to larger $\gamma$ values. We show this in Figure.~\ref{fig: caus com eigs}.

\subsection{Extended stability analysis: complex $\gamma$'s and particle-hole antisymmetric sector}
\label{app: stability}

In the main text, we displayed results of the linear stability analysis on the real $\gamma$ axis, and only in the $E-$antisymmetric (particle-hole symmetric) sector. Here we extend the analysis to include complex $\gamma$ and the particle-hole antisymmetric sector. 

In the following calculations we set $T=1$ and use anti-symmetrized and symmetrized energy grids with $E_{\rm max}=15$, $dE=0.05$.
We use a band of phonons with $\Delta\Omega_0/\Omega_0=0.6$. Specifically we use a set of $N_\Omega=(2\Delta \Omega_0/dE+1)$ equally spaced phonon frequencies in the range $\Omega_0-\Delta\Omega_0\leq\Omega_i\leq \Omega_0+\Delta\Omega$, all with equal couplings $g_i = \frac{g}{\sqrt{N_\Omega}}$ (the exact value of $\Delta \Omega_0$ and form of the phonon band do not alter results significantly).

\begin{figure}[t]
\centering
\includegraphics[width=0.9\textwidth]{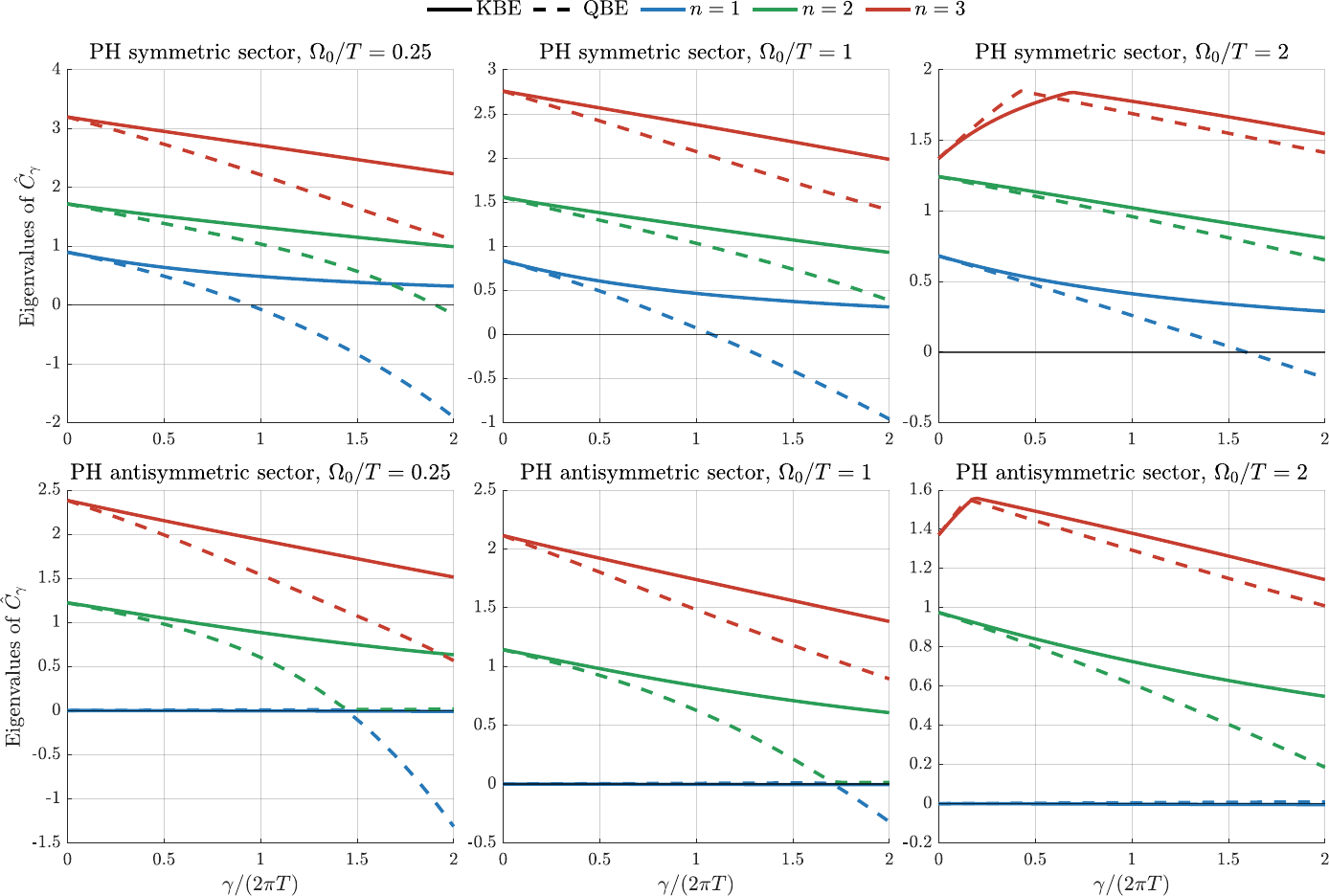} 
\caption{Comparison of three lowest lying eigenvalues of $\hat{C}_\gamma$ using KBE (solid) and QBE (dashed) in the particle-hole symmetric (top) and antisymmetric (bottom) sectors. The antisymmetric sector always contains a zero mode corresponding to the conserved electron density.}
\label{fig: ph}
\end{figure}

In Figure~\ref{fig: ph} we repeat the same analysis applied to Fgiure.~\ref{figure 1}(top) of the main text, including the particle-hole antisymmetric sector. Namely, we calculate the spectrum of $\hat{C}_\gamma$ for positive $\gamma>0$ (so that rates $\gamma_*$ are those that satisfy the equation $\gamma_*\in -\tfrac{g^2}{T} {\rm spec}(\hat{C}_{\gamma_*})$) and plot the 3 lowest lying eigenvalues in both sectors. Note that for both QBE and KBE, there is a zero eigenvalue independent of $\gamma$, which at $\gamma=0$ corresponds to the conserved electron density. Excluding this mode, all eigenvalues of the KBE operator at positive, real, $\gamma$ are always strictly positive, and thus no positive $\gamma_*$ is a solution for a rate of the KBE kinetic operator.

We now extend the stability analysis to complex values of $\gamma$ with positive real part. 
We are searching for points that satisfy:
\begin{equation}
    \gamma = -\frac{g^2}{T} \lambda(\gamma) .
    \label{gamma eig relation}
\end{equation}
for some eigenvalue $\lambda(\gamma)$ of the collision operator $\hat{C}_\gamma$. To identify such points efficiently, we define
\begin{equation}
    \theta(\gamma) = \min_{\lambda(\gamma)\in {\rm spec}(\hat{C}_\gamma)} \left|\arg\left(\frac{\lambda(\gamma)}{-\gamma}\right)\right|,
\end{equation}
such that values of \(\gamma\) for which \(\theta(\gamma)=0\) are in one-to-one correspondence with solutions of Eq.~\eqref{gamma eig relation} for some value of \(g^2/T\).

\begin{figure}[t]
\centering
\includegraphics[width=0.9\textwidth]{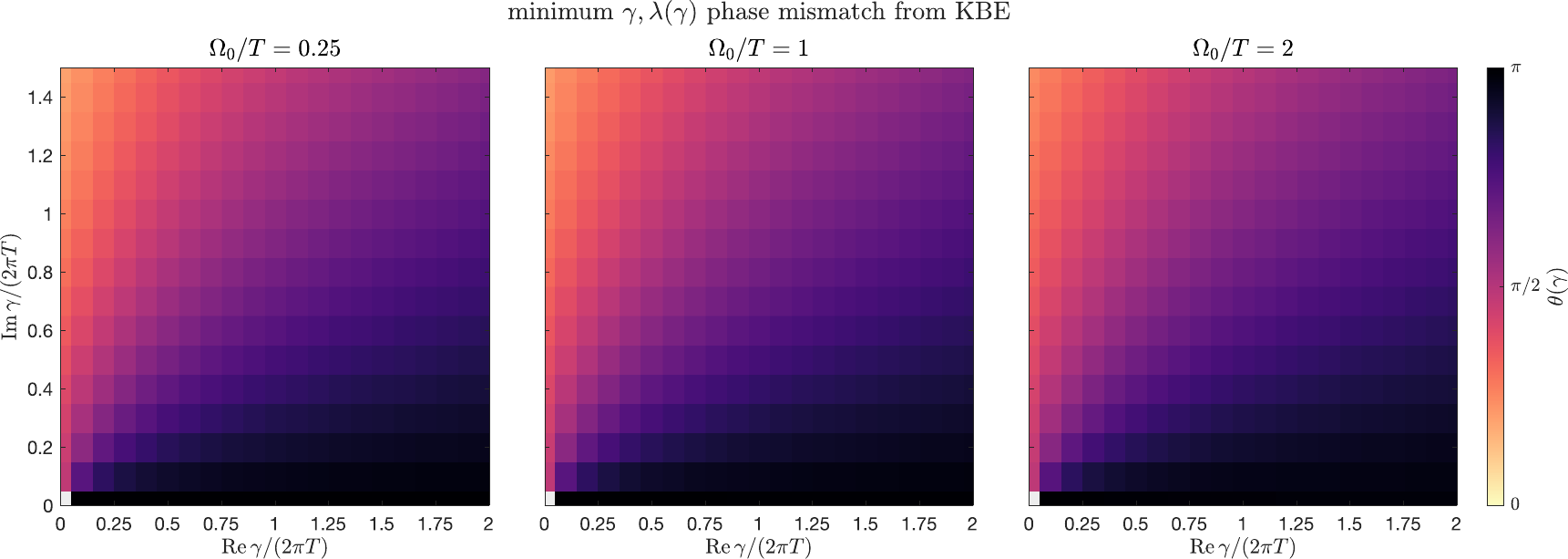} 
\caption{Minimal phase mismatch between $\gamma$ and the spectrum of $\hat{C}_\gamma$, denoted $\theta(\gamma)$, using the KBE. }
\label{fig: phase}
\end{figure}

In Fig.~\ref{fig: phase} we plot \(\theta(\gamma)\) over a broad region of the complex plane with \({\rm Re}\,\gamma,{\rm Im}\gamma>0\)  corresponding to frequencies in the upper half-plane, for several values of \(\Omega_0/T\) (results at ${\rm Im}\gamma<0$ are identical; since we are dealing with Hermitian dynamics, all solutions $\gamma$ must appear in complex-conjugate pairs). This includes both the particle-hole symmetric and antisymmetric sectors, but excludes the branch of zero eigenvalues which is connected to the density-conserved zero mode. The absence of zeros of $\theta(\gamma)$ anywhere on this complex half plane shows that \(\hat{K}_\gamma\) has no zero modes in this region, and thus the KBE has no unstable modes.

\subsection{Derivation of linearized sources and observables}
\label{app: linearization}

Derivation of the sources is similar to the derivation of the linearized KBE itself. 
As a warm-up, consider the simpler case of a time-dependent electron-phonon coupling: $g(t)=(1+\epsilon_\omega(t))g$. In this case, the self energies are (for $X=R/A/K$)
\begin{equation}
    \Sigma^X(t_1,t_2) = \frac{g(t_1)g(t_2)}{g^2} \Sigma^X[g^K],
\end{equation}
where $\Sigma^X[g^K]$ is the same self-energy functional defined in the End Matter. 
Expanding this to linear order in $\epsilon_\omega$, using $\Sigma^X[g^K]=\Sigma^X_{\rm eq}+\epsilon_\omega D^X\delta g^K$,
\begin{align}
    \Sigma^X(t_1,t_2) &= \left(1+\epsilon_\omega(t_1)+\epsilon_\omega(t_2)\right)\left(\Sigma^X_{\rm eq}(\tau) + \epsilon_{\omega}\left(t\right) D^X(\tau)\delta g^K(\tau)\right)
    \\ \nonumber  
    &= \Sigma^X_{\rm eq}(\tau) + \epsilon_\omega(t) \delta \Sigma^X(\tau).
\end{align}
We split $\delta \Sigma^X$ into two contributions: the ``direct'' contribution from the drive (i.e. $g(t_{1,2})$) which we call $\delta_0\Sigma^X$, and the electron response (i.e. $\delta g^K$). Wigner transforming, these are:
\begin{align}
    \delta \Sigma^{X}(E)&= \hat{D}^X \delta g^K + \delta _0\Sigma^X, \\ \nonumber 
    \delta_0\Sigma^X & = \Sigma^X_+ + \Sigma^X_-.
\end{align}
where, as a reminder, $A_\pm \equiv A_{\rm eq}(E\pm\omega/2)$.
Plugging this into the quasiclassical KBE and splitting these two terms, the $\hat{D}^X$'s give the scattering terms in the kinetic operator $\hat{K}_\omega$ (as defined in the End Matter), while the ``direct'' perturbations $\delta_0 \Sigma^X$ can be combined into a source term on the RHS:
\begin{align}
    S_\omega(E) &= i\left(2\delta_0 \Sigma^K(E)  -g^K_- \delta_0\Sigma^R(E) + g^K_+\delta_0\Sigma^A(E)\right).
\end{align}
In this example of a drive in the interaction strength, the source is
\begin{align}
S_\omega(E)=i(\Sigma^R_+ + \Sigma^A_-)(g^K_+-g^K_-).
\end{align}
% For a drive of the phonon temperature as a function of the COM time $T\to (1+\epsilon_\omega(t))T$
% \begin{align}
%     i\delta \Sigma^K(E)&=\frac{\eta}{2} \left(g^{K}_{\rm eq}\left(E-\Omega_{0}\right)+g_{\rm eq}^{K}\left(E+\Omega_{0}\right)\right) \\ \nonumber 
%     \delta \Sigma^{R/A}(E) &= \mp i \eta 
% \end{align}
% with $\eta \equiv\frac{\pi g^2}{4T\sinh^2\left(\frac{\Omega_0}{2T}\right)}$ gives the source 
% \begin{equation}
%     S_\omega = \eta\left(g^{K}_{\rm eq}\left(E-\Omega_{0}\right)+g_{\rm eq}^{K}\left(E+\Omega_{0}\right)-g^K_+-g^K_-\right).
% \end{equation}

We now turn to the slightly more complicated issue of a drive that couples to the electron dispersion, which cannot be inserted straightaway into the quasiclassical equations since momentum has already been integrated over. We consider a drive of an operator of the form:
\begin{align}
    H&=H_{\rm el-ph}+\epsilon_\omega(t)\hat{h} \\ \nonumber 
    \hat{h} &= \sum_k h(\xi_k)c^\dagger_k c_k 
\end{align}
The function $h(\xi_k)$ depends on momentum only through the QP energy $\xi_k$. To incorporate it, we must insert it into the Schwinger-Dyson-Keldysh (SDK) equations of the unprojected Green's functions\cite{rammer1986quantum}:
\begin{align}
    \left[G_0^{-1}\check{I}-\check{\Sigma},\check{G}\right]_\circ=0 \label{SDK_mat}
\end{align}
where $\check{A}=\begin{pmatrix}
A^R & A^K \\ \nonumber 
0 & A^A
\end{pmatrix}$, $\check{I}$ is the identity, and $[\cdot,\cdot]_\circ$ is a commutator with respect to both Moyal product and  $\check{\cdot}$ matrix multiplication structure (the equation is diagonal in momentum $k$). The bare Green's function is \begin{equation}G_0^{-1}=E-\xi_k - \epsilon_\omega(t)h(\xi_k).\end{equation}
When written in terms of components, Eq.~\eqref{SDK_mat} corresponds to
\begin{align}
    \left[G_0^{-1}-\Sigma^{R/A},G^{R/A}\right]_\circ&=0  \label{SDK_comps}\\ \nonumber  
    \left[G_0^{-1},G^{K}\right]_\circ&=\Sigma^R\circ G^K-G^K\circ \Sigma^A+\Sigma^K\circ G^A-G^R\circ \Sigma^K .
\end{align}
The perturbation $\hat{h}$ will also perturb $G^{R/A}$ such that the projected $g^{R/A}$ are no longer $\pm 1$. Explicitly, the perturbations $\delta G^{R/A}$ will obey:
\begin{align}
    [E-\epsilon_\omega(t)h(\xi_k)-\Sigma^{R/A}_{\rm eq}-\epsilon_\omega(t)\delta \Sigma^{R/A}, G_{eq}^{R/A}+\epsilon_\omega(t)\delta G^{R/A}]_\circ =  0
\end{align}
We can now integrate this equation $\frac{i}{\pi}\int d\xi $ to obtain an equation for $\delta g^{R/A}$:
\begin{equation}
    [E-\Sigma^{R/A}_{\rm eq}, \epsilon_\omega \delta g^{R/A}]_\circ-[ \epsilon_\omega,h^{R/A}_{\rm eq}]_\circ -[ \epsilon_\omega \delta \Sigma^{R/A},g^{R/A}_{\rm eq}]_\circ  = 0 
\end{equation}
where we have defined:
\begin{equation}
    h^{R/A}_{\rm eq}(E) \equiv \frac{i}{\pi} \int d\xi h(\xi)G^{R/A}_{\rm eq}(E,\xi).
\end{equation}

Since $g^{R/A}_{\rm eq}=\pm 1$ the commutator with the $\delta \Sigma^{R/A}$ term is $0$ and at  the linearized level we obtain a closed,  diagonal equation for $\delta g^{R/A}$ which does not depend on the Keldysh component $\delta g^K$: 
\begin{align}
    \delta g^{R/A} & = -\frac{h^{R/A}_+-h^{R/A}_-}{\omega -(\Sigma^{R/A}_+-\Sigma^{R/A}_-)} ,
\end{align}
having used
\begin{equation}
    [A,\epsilon_\omega]_\circ = (A_+-A_-)\epsilon_\omega.
\end{equation}

We can now derive the source in the equation for $g^K$, by projecting the equation for $G^K$ in Eq.~\eqref{SDK_comps} in the same manner. This will include contributions from (1) the $h(\xi_k)$ term in $G_0^{-1}$, (2) the corrections $\delta G^{R/A}$ to the explicit $G^{R/A}$ terms, and (3) the self energies, which also depend on $\delta g^{R/A}$:
\begin{align}
    \delta_0 \Sigma^{R/A}&=\hat{D}^K\delta g^{R/A},\\ \nonumber 
    i\delta_0 \Sigma^K &= \hat{\mu} \left(\delta g^R -\delta g^A\right).
\end{align}
Collecting all of these and projecting $\frac{i}{\pi}\int d\xi$ (and multiplying by $-i$ for the convention used in the main text) gives the source: 
\begin{equation}
    S_\omega = i\left(2\delta_0 \Sigma^K(E)  -g^K_- \delta_0\Sigma^R(E) + g^K_+\delta_0\Sigma^A(E)+\Sigma^K_- \delta g^R-\Sigma^K_+ \delta g^A +h^K_+-h^K_-\right)
    \end{equation}
    with $h^K_{\rm eq}=\tanh\left(\tfrac{E}{2T}\right)\left(h^R_{\rm eq}-h^A_{\rm eq}\right)$. 

    We now turn to developing a formula for the response of single-body electron observables in terms of the electron's response $\delta g^K$. We define a single body observable similar to $\hat{h}$
\begin{equation}
    \hat{b}=\sum_k b(\xi_k)c^\dagger_k c_k .
\end{equation}
In the presence of a time-dependent drive, the expectation value is given by the lesser Green's function:
\begin{equation}
    \left<\hat{b}(t)\right> = -i \int_k \int \frac{dE}{2\pi} b(\xi_k) G^<(t,E,\xi_k)
\end{equation}
Using $-G^<=\frac{1}{2}(G^R-G^A-G^K)$ we can write the perturbation in this observable as:
\begin{equation}
    \frac{\delta\left<\hat{b}(t)\right>}{\epsilon_\omega(t) }= \frac{\nu_F}{4}\int dE \left(\delta b^R(E)-\delta b^A(E)-\delta b^K(E) \right)
\end{equation}
with $\delta b^X\equiv \frac{i}{\pi}\int d\xi b(\xi)\delta G^X(E,\xi)$. Since we now know $\delta g^K$ we can calculate the full perturbations to the self energies:
\begin{equation}\delta\Sigma^X=\hat{D}^X\delta g^K + \delta_0\Sigma^X,
\end{equation}
and thus don't need to invert any non-diagonal kinetic operator again.
The equations of motion for the $\delta b$'s are found by projecting the SDK equations Eq.~\eqref{SDK_comps} with $\int d\xi b(\xi)$,
\begin{align}
    \delta b^{R/A}   \ &= -\frac{[ hb^{R/A}_{\rm eq},\epsilon_\omega]_\circ+[ b^{R/A}_{\rm eq},\epsilon_\omega]_\circ\delta\Sigma^{R/A}}{[E-\Sigma^{R/A}_{\rm eq},\epsilon_\omega]_\circ} = -\frac{hb^{R/A}_+-hb_-^{R/A}+(b^{R/A}_+-b_-^{R/A})\delta \Sigma^{R/A}}{\omega -(\Sigma^{R/A}_+-\Sigma^{R/A}_-)} 
\end{align}
$b_{\rm eq}^X$ is defined as $h^X_{\rm eq}$, and in the presence of an $\hat{h}$ drive $hb_{\rm eq}^X=\frac{i}{\pi}\int d\xi h(\xi)b(\xi) G^X_{\rm eq}(E,\xi)$ (else that term should be omitted). Finally, the Keldysh component:
\begin{equation}
    \delta b^K = \frac{b^K_-\delta\Sigma^R - b^K_+\delta\Sigma^A-(b^R_+-b^A_-)\delta\Sigma^K+\Sigma^K_+ \delta b^A - \Sigma^K_- \delta b^R+hb^K_--hb^K_+}{\omega -\left(\Sigma^R_+-\Sigma^A_-\right)}.\label{dbK}
\end{equation}
Note that Eq.~\eqref{dbK} does \textit{not} have a well-behaved $\Omega_0/T\to 0$ limit like the kinetic operator $\hat{K}_\omega$, in the sense that the dependence on the quasiclassical scattering rate $\Gamma_0 \sim \lambda$, which diverges in this limit, does not cancel out. The $\Sigma^K_\pm$ in the numerator, as well as the $\Sigma^R_+-\Sigma^A_-$ in the denominator both depend explicitly on $\Gamma_0$ and thus dominate both numerator and denominator in this limit. In contrast, in the equation for $\delta g^K$ these terms both combine in $\hat{K}_\omega$ and give the diffusion operator $\partial_E^2$.
This distinction stems from the fact that $\delta g^K$ is the $0$th spectral moment, so does not depend on the broadened spectral weight in the presence of $\Gamma_0$, unlike $\delta b^K$ which is very sensitive to it.

To summarize, the calculation order is  roughly
\begin{equation}
\delta g^{R/A}\to \delta g^K \to \delta\Sigma^X\to \delta b^{R/A}\to \delta b^K \to \delta \left<\hat{b} \right>,
\end{equation}
and the only step in this calculation where a large operator is inverted is the 2nd step 
$\delta g^K=\hat{K}_\omega^{-1}S_\omega$.

In the main text we make use of the Lorentzian-weighted kinetic energy:
\begin{equation}
     h(\xi)=b(\xi)=\frac{\xi}{1+\xi^2/\Lambda^2}
\end{equation}
for which 
\begin{align}
    h^{R/A}_{\rm eq}(E) =b^{R/A}_{\rm eq}(E) &=  \frac{\pm \Lambda^2}{Z^{R/A}\pm i\Lambda } \\ \nonumber 
    hb^{R/A}_{\rm eq}(E) &= \frac{i\Lambda^3Z^{R/A}}{2\left(Z^{R/A}\pm i\Lambda\right)^2}
\end{align}
where $Z^{R/A}(E)=E-\Sigma^{R/A}_{\rm eq}(E)$.
A possible probe for the particle-hole antisymmetric sector is the Lorentzian-weighted density:
\begin{equation}
     h(\xi)=b(\xi)=\frac{1}{1+\xi^2/\Lambda^2}
\end{equation}
for which
\begin{align}
    h^{R/A}_{\rm eq}(E) =b^{R/A}_{\rm eq}(E) &= \frac{i\Lambda}{ Z^{R/A}\pm i\Lambda} \\ \nonumber 
    hb^{R/A}_{\rm eq}(E) &=\frac{
i\Lambda\left(Z^{R/A}\pm 2i\Lambda\right)
}{
2\left( Z^{R/A}\pm i\Lambda\right)^2
}
\end{align}

Finally, for the QBE version of these calculations, we simply replace all $A_\pm \to (1\pm\frac{\omega}{2}\partial_E)A_{\rm eq}$ in all above formulas.

For completeness, we detail here the procedure behind the calculations of $\chi^R_{\rm kin-kin}(\omega)$ in the main text (Figure~\ref{figure 1} bottom) and extend them to show the $\lambda$ dependence. 
\subsubsection*{Numerical calculation}
The KBE requires shifting operators on the energy axis by $\pm \omega/2$, so that reaching large frequencies requires a very wide energy axis. Since the real part of the response $\delta g^K(E)$ has slowly decaying tails in $E$, we use the following extrapolation scheme. For every value of $\omega$, we perform the calculations described in Sec.~\ref{app: linearization} on energy grids with $E_{\rm max} = E_{\rm max,0} + \omega/2$ using $E_{\rm max,0}/(g^2/T) = 30,40$ and then extrapolate the resulting $\chi^R(\omega,E_{\rm max,0})$ linearly in $1/E_{\rm max,0}\to 0$, pointwise in $\omega$. This allows us to calculate $\chi^R(\omega)$ over a finite range of $\omega$, which we extend up to $\omega\leq 40 g^2/T$. The QBE does not involve $\pm \omega/2$ energy shifts, so we can use a fixed energy axis. We extrapolate the QBE calculations using a finite range of $E_{\rm max}\leq 100$ independent of frequency, which allows calculations up to arbitrarily large $\omega$. All calculations are performed using $dE/T=0.2$ and a band of phonons with frequencies $ \Omega_i/T\in \{0.4,0.6,0.8,\cdots,1.6\}$, all coupled equally with $g_i = g/\sqrt{7}$.

To apply Kramers-Kronig relations on the real parts, we fit their high frequency tails to

\begin{align}
    {\rm Re}\chi^R(\omega)_{\rm kin- kin} \propto \begin{cases} A+B/\omega^2 & {\rm QBE} \\ C/|\omega| & {\rm KBE} \end{cases}
\end{align}
Note the differences in the high$-\omega$ forms of the two:  the QBE approaches a constant at high frequencies since the QBE source grows like $S_\omega\propto i\omega $. On the other hand, the KBE decays like $1/|\omega|$ - this peculiar form is due to the Lorentzian weight in our definition of $h(\xi)$, and this behavior agrees with that of the Kubo-formula calculation of the appropriate bubble diagram at high $\omega$ (where vertex corrections can be ignored).

These tails are then used to complete the high frequency part of the KK integrals. Throughout the whole process, we verify that the $E_{\rm  max,0}$ extrapolation converges well, and that the $\omega$ range is wide enough so that these high frequency tails fit well.

\begin{figure}[t]
\centering
\includegraphics[width=0.7\textwidth]{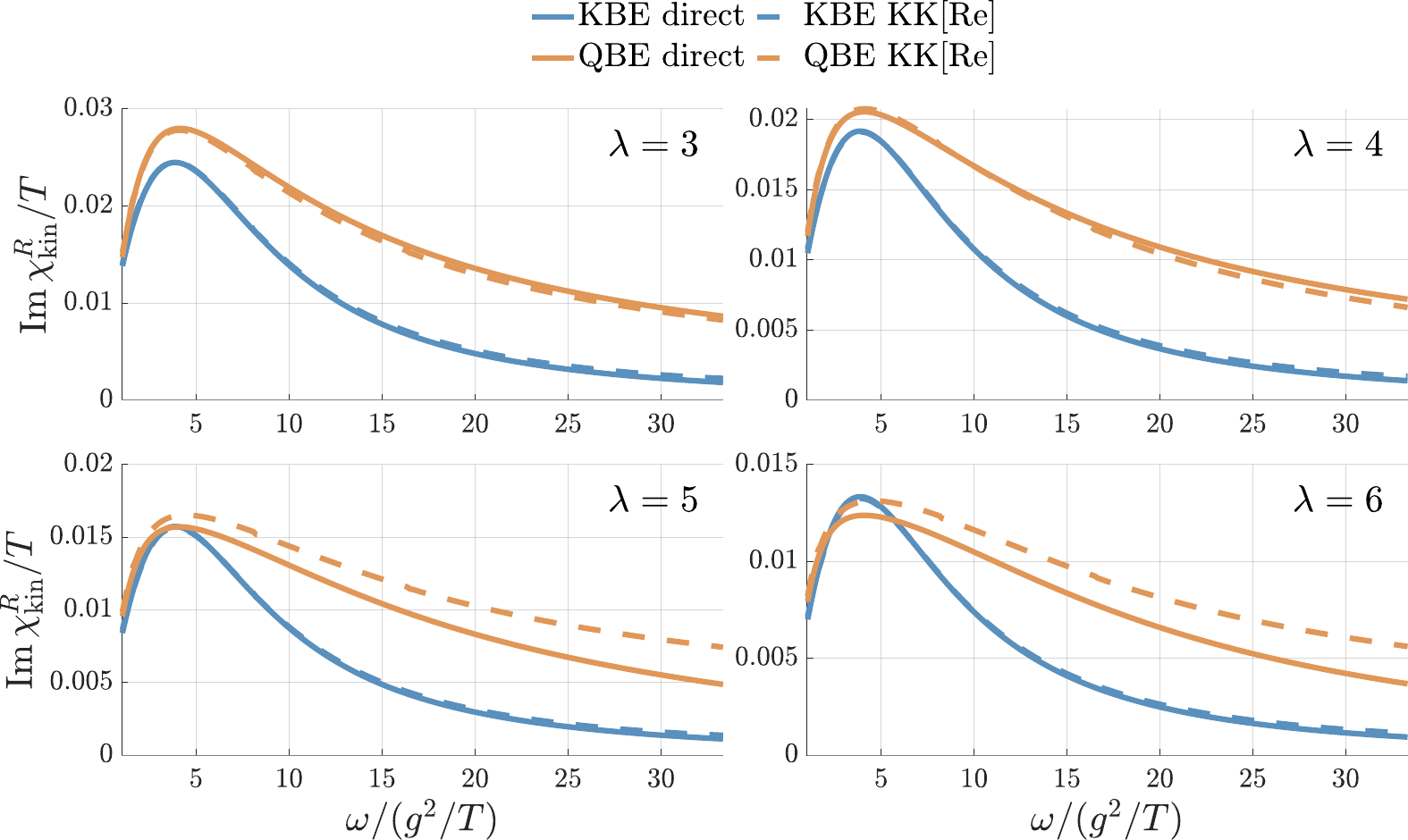} 
\caption{Real frequency susceptibility to the Lorentzian-weighted kinetic energy with $T=1$, comparing direct calculations of its imaginary part (solid) with indirect calculations using the KK relations (dashed). For $\lambda>\lambda_c\approx 3.3$, the two disagree in the QBE response, indicating the presence of a pole in the UHP. The plots have been truncated at $\omega/(g^2/T) \geq 1$ since at smaller frequencies there are fine features which require finer frequency resolution to reproduce using KK.}
\label{fig: chi appendix}
\end{figure}

We show the results of this calculation, for fixed $\Omega_0=T$ and varying $\lambda = 3,4,5,6$ in Figure~\ref{fig: chi appendix}. For these values of $\Omega_0,T$, the instability in the QBE occurs around $\lambda_c\approx 3.3$. This corresponds to the point where the KK transform of the QBE response function ceases hold.

% A few convenient examples of $h^{R/A}_{\rm eq}$ are the regular and Lorentzian-weighted kinetic energies:
% \begin{align}
%     h(\xi)=\xi\qquad\qquad\space &\Rightarrow h^{R/A}_{\rm eq}(E) = -i {\rm const} \pm \left(E-\Sigma^{R/A}_{\rm eq}\right) \\ \nonumber 
%     h(\xi)=\frac{\xi}{1+\xi^2/\Lambda^2} &\Rightarrow h^{R/A}_{\rm eq}(E) = \frac{\pm \Lambda^2}{i\Lambda\pm \left(E-\Sigma^{R/A}_{\rm eq}\right)}
% \end{align}

\subsection{Details on relaxation rate calculation}
\label{app: relaxation}
\subsubsection*{Characterizing behavior at small, real $\gamma$} 
\subsubsection{$\Omega_0/T\to 0$}
We begin by studying the limit $\Omega_0/T\to 0$ in which  $\hat{C}_\gamma$ becomes a continuous linear operator which only depends explicitly on $T$.  At $\gamma=0$, $\hat{C}_{\gamma=0}$ is equivalent to the linearized collision operator of the Boltzmann equation, and is given by the form
\begin{equation}
\hat{C}_0=-\pi T\left(T\partial_E^2 +\partial_E \tanh(E/2T)\right).
\end{equation}
with $\partial_E$ acting on everything to the right. It is easy to verify that $\hat{C}_0$ has a right zero mode 
\begin{equation}
    \delta g^K(E)=\frac{1}{\cosh^2\left(\frac{E}{2T}\right)} \sim \frac{\partial g^K_{\rm eq}}{\partial E}
    \label{zero mode}
\end{equation} which corresponds to a shift in the electron density.
This operator can be related to the exactly solvable ``Pöschl-Teller'' Hamiltonian \cite{poschl1933bemerkungen} by a similarity transformation:
\begin{align}
    \hat{C}_0 &= \frac{\pi}{4} (\cosh(x))^{-1} \hat{H}  \cosh(x), \\ \nonumber  
    \hat{H} &=  \mathcal{A}^\dagger\mathcal{A}, \quad  \mathcal{A} = \partial_x + \tanh(x),
\end{align}
where $x=\frac{E}{2T}$.

Defining the ``twin'' operator $\check{H}=\mathcal{A}\mathcal{A}^\dagger = -\partial_x^2 + 1$, we can construct eigenfunctions of $\hat{H}$ using those of $\check{H}$:
\begin{align}
    \check{H}\ket{p}=\lambda_p\ket{p}\Rightarrow \hat{H}\mathcal{A^\dagger}\ket{p}=\mathcal{A^\dagger}\check{H}\ket{p }=\lambda_p \mathcal{A^\dagger}\ket{p}.
\end{align}
Since $\ket{p}$ are plane waves, the eigenfunctions of $\hat{H}$ (excluding the zero mode)  are \begin{equation}
\ket{\psi_p}\propto\mathcal{A^\dagger}\ket{p} = (ip+\tanh(x))e^{-ipx}.
\end{equation} 
(up to a normalization constant).

Thus, the spectrum of $\hat{C}_0$ is $\left\{0,\lambda_p= \frac{\pi }{4}(1+ p^2)\right\}$ where $p$ is some continuous label for momenta on the $x$ axis.

Consider now small values of $|\gamma|\ll T$ as a perturbation to the operator, i.e. $\hat{C}_{\gamma} \approx \hat{C}_{0} + \gamma \partial_\gamma \hat{C}_{0}+\cdots$. Since the  spectrum of $\hat{H}$ is comprised of dispersive modes delocalized on the energy axis, the perturbation $\partial_\gamma \hat{C}_{0}$ (which decays at large $E$) cannot affect the dispersion $\lambda_p$. Its only possible effect is to nucleate localized bound states below the continuum threshold, which only happens if the perturbation is sufficiently attractive. For negative $\gamma$'s this is not the case, and in the absence of such bound states the continuum eigenvalues remain independent of $\gamma$. Thus the asymptotic relaxation rate, defined as the least negative, nonzero $\gamma$ which satisfies Eq.~\eqref{eig condition} (excluding the density zero mode), is given exactly by the value of the continuum threshold $-\gamma_0=\frac{g^2} {T}\lambda_{p=0}=\frac{\pi  g^2}{4T}$ for a nonzero  range of $g$'s, i.e. the ``naive'' perturbative result extends non-perturbatively in $g$.
We also note that since the gap is not to a single isolated mode, but rather to a continuum with DOS $\sim 1/\sqrt{\lambda_p-\lambda_0}$, the asymptotic behaviour acquires a power law correction $\phi(t\to\infty) \sim e^{\gamma_0 t}/\sqrt{t}$,

\subsubsection{Relation to Allen's formula}

As an aside, we show that the above form of the collision operator reproduces the results of \cite{allen1987theory}.

In the case where there are additional strong electron-electron interactions, we may assume that the electrons are thermally equilibrated with each other at a thermal distribution whose temperature is different than the phonons'. This means that the only possible perturbation is
\begin{equation}
    \delta g^K \sim \partial_T g^K_{\rm eq} \propto \frac{x}{\cosh^2{x}} \equiv (\cosh(x))^{-1}  \ket{\psi_T}.
\end{equation}

To span it we must normalize the eigenmodes of $\ket{\psi_p}$. The normalization is  determined by:
\begin{align}
    \braket{\psi_p|\psi_{p'}}=N_p^2 \braket{p\left|\mathcal{A}\mathcal{A}^\dagger\right|p'}=N_p^2(p^2+1)\braket{p|p'}\Rightarrow N_p = \frac{1}{\sqrt{1+p^2}},
\end{align}
and thus we can span
\begin{equation}
    \hat{H} = \int \frac{dp}{2\pi}  (p^2+1) \ket{\psi_p}\bra{\psi_p}
\end{equation}

Then the projected collision operator is just
\begin{equation}
    \frac{g^2}{T}\hat{C}_{\rm proj} =  -\gamma_{\rm proj}(\cosh(x))^{-1} \ket{\psi_T}\bra{\psi_T}  \cosh(x)
\end{equation}
and the projected relaxation rate is given by
\begin{align}
    -\gamma_{\rm proj} = \frac{\pi g^2 }{4T}\frac{\braket{\psi_T|\hat{H}|\psi_T}}{\braket{\psi_T|\psi_T}}.
\end{align}
The relevant overlaps are
\begin{align}
    \braket{\psi_T|\psi_T} &= \int dx \frac{x^2}{\cosh^2x} = \frac{\pi^2}{6} \\ \nonumber 
    \braket{\psi_p|\psi_T}&= \frac{1}{\sqrt{1+p^2}}\int dx e^{ipx}\left[\mathcal{A} \frac{x}{\cosh x}\right] \\ \nonumber  &=\frac{1}{\sqrt{1+p^2}} \int dx e^{ipx}\frac{1}{\cosh x} = \frac{\pi}{\sqrt{1+p^2}} \frac{1}{\cosh\left(\frac{\pi p}{2}\right)} \\ \nonumber  
    \Rightarrow \braket{\psi_T|\hat{H}|\psi_T} & = \int \frac{dp}{2\pi} (1+p^2) \left| \braket{\psi_p|\psi_T}\right|^2 = \frac{\pi}{2} \int \frac{dp}{\cosh^2\left(\frac{\pi p}{2}\right)}  = 2
\end{align}
which results in 
\begin{equation}
   - \gamma_{\rm proj} = \frac{3g^2}{\pi T}
\end{equation}
as first calculated in \cite{allen1987theory}.

\subsubsection{Finite $\Omega_0/T$}

We now consider the spectrum at finite $\Omega_0/T$. Obviously, when $\Omega_0\gg T$ relaxation is exponentially suppressed, and the resulting eigenvalues are $\sim e^{-\Omega_0/T}$. We instead focus on intermediate values of $\Omega_0/T$. At $\gamma=0$ and for a sharp phonon at a single frequency, inelastic phonon scattering only couples points on the energy axis separated by integer multiples of $\Omega_0$. This leads to a large degeneracy for all solutions found in the above used continuum approximation $\Omega_0/T\to0$, i.e. every eigenmode of $\hat{C}_0$ can be split into degenerate sub-modes 

\begin{equation}\delta g^K(E)\to \delta g^{K,( n)}(E)=e^{iq_nE}\delta g^K(E),\end{equation}
 with $q_n=\frac{2\pi n}{\Omega_0}$ and $n\in\mathbb{Z}$ .  Specifically, the bound state corresponding to the density-shift zero mode in Eq.~\eqref{zero mode} also splits into degenerate modes, creating an infinite number of zero modes at $\gamma=0$ (orthogonal to the $n=0$ density-shift mode).

Introducing a finite phonon bandwidth $\Delta \Omega_0$ (i.e. spreading the interaction between multiple phonon modes with frequencies in a range $\sim \Delta \Omega_0$ around $\Omega_0$) allows for diffusion below $\Omega_0$, which leads to a splitting of these sub-modes.  Defining $r=\Delta \Omega_0/\Omega_0<1$, the splitting can be estimated to be of the order $D q_n ^2$ with diffusion constant 
\begin{equation}
    D\sim r^2 T\Omega_0 \coth({\Omega_0/2T}).
\end{equation}
 Out of these, the lowest lying nonzero modes are the $n=\pm 1$ submodes of the density-shift mode, which have $\lambda^{(1)}(\gamma=0)\sim D/\Omega_0^2$. When $\Omega_0/T$ is sufficiently small this eigenvalue is well above the continuum threshold, but eventually descends below it as $\Omega_0$ becomes a significant fraction of $T$ (i.e. $\Omega_0/T\sim r$).

The important distinction between these modes and the continuum modes discussed earlier (energy plane-waves) is that this is a \textit{localized bound state} on the energy axis, and thus is not impervious to the perturbation introduced by a small $\gamma$. This causes a linear drift of its eigenvalue with $\gamma$, as\footnote{The slope with respect to $\gamma$ must be positive; if it were negative, the eigenvalues would have become negative at positive $\gamma$, implying an instability. This can be verified easily  by a direct calculation of the matrix elements of this mode with $\partial_\gamma \hat{C}_{0}$. } $\lambda^{(1)}(\gamma<0)- \lambda^{(1)}(0)\propto -|\gamma|/T$, which inevitably leads to it crossing zero $\lambda^{(1)}(\gamma=\gamma_*)=0$ at \begin{equation}
    -\gamma_* \propto T\lambda^{(1)}(0) \propto \frac{r^2T^2}{\Omega_0}\coth(\Omega_0/2T). \label{gamma star}
\end{equation} This  zero creates a bottleneck for relaxation, and as $g/T\to \infty$ this branch of solutions saturates at $\gamma(g/T\to \infty) \to \gamma_*$. Note that this is a spectral effect, absent in the perturbative Boltzmann equation (requiring the first $\gamma$ derivative).

\begin{figure}[t]
\centering
\includegraphics[width=0.7\textwidth]{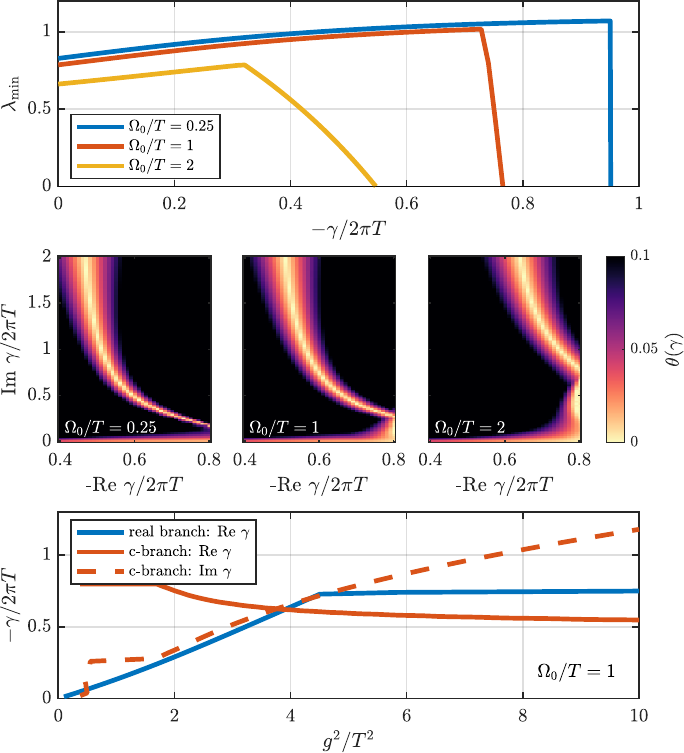} 
\caption{Analysis of relaxation modes using the KBE. \textbf{Top:} Minimal eigenvalue of $\hat{C}_\gamma$ at negative, real, $\gamma$ for various $\Omega_0/T$ values. The small, nonzero slope around $\gamma=0$ is due to the finite extent of the energy grid, $E_{\rm max}=15$, and flattens out as larger $E_{\rm max}$ values are used. \textbf{Center:} Plots of the minimal angle $\theta(\gamma)$ in the complex $\gamma$ plane, for various $\Omega_0/T$ values. Eigenmodes of the KBE correspond to points satisfying $\theta(\gamma)=0$. \textbf{Bottom:} Example of the competition between the real and complex branches for $\Omega_0/T=1$. The complex branch takes over around $g^2/T^2\gtrsim 4$. }
\label{figure 2}
\end{figure}

\subsubsection*{Larger $\gamma$ and complex $\gamma$}

Extending to larger as well as complex values of $\gamma$ requires numerical treatment. As in Appendix~\ref{app: stability}, in the following calculations we set $T=1$ and use an anti-symmetrized energy grid with $E_{\rm max}=15$, $dE=0.05$.
We use a band of phonons with $\Delta\Omega_0/\Omega_0=0.6$. Specifically we use a set of $N_\Omega=(2\Delta \Omega_0/dE+1)$ equally spaced phonon frequencies in the range $\Omega_0-\Delta\Omega_0\leq\Omega_i\leq \Omega_0+\Delta\Omega$, all with equal couplings $g_i = \frac{g}{\sqrt{N_\Omega}}$ (the exact value of $\Delta \Omega_0$ and form of the phonon band do not alter results significantly).

In Fig.~\ref{figure 2}(top) we plot the minimal eigenvalue $\lambda_{\rm min}(\gamma)$ at negative values of $\gamma$ for various values of $\Omega_0/T$. As expected, at $\gamma\to0,\Omega_0\to0$ the eigenvalue approaches $\pi/4$. While it should be constant in a range of $\gamma$'s, it disperses weakly with $\gamma$ due to finite size effects.
For small $\Omega_0/T$ it appears like some of the eigenvalues diverge $\lambda(\gamma)\to -\infty$ at $\gamma\to -2\pi T$, although this might be an artifact related to the singularity in the collision operator at ${\rm Re}\gamma=-2\pi T$ (In this analysis, since this suspicious eigenvalue is never the slowest relaxing mode we do not give this more thought. Fixing this would require writing down the ``anti-causal'' version of the collision operator, as explained in Appendix~\ref{app: mer cont}). For large enough $\Omega_0/T$ there is a physically meaningful (since it does not occur close to the singularity ${\rm Re}\gamma=-2\pi T$) bound state with eigenvalue crossing $0$ at $\gamma_*$ as described by Eq.~\eqref{gamma star}.

We also extend this analysis to the complex $\gamma$ plane. To simplify the search for solutions to Eq.~\eqref{eig condition}, we define the angle 
\begin{equation}
    \theta(\gamma) = \min_{\lambda(\gamma)\in {\rm spec}(\hat{C}_\gamma)} \left|\arg\left(\frac{\lambda(\gamma)}{-\gamma}\right)\right|,
\end{equation}
such that solutions to $\gamma = -\frac{g^2}{T} \lambda(\gamma)$ are possible for some value of $g/T$ when $\theta(\gamma)=0$, i.e. when an eigenvalue is antiparallel to $\gamma$. 
In Fig.~\ref{figure 2}(center) we plot $\theta(\gamma)$ throughout the complex $\gamma$ plane for the same three values of $\Omega_0/T$. For low values of $\Omega_0/T$ there is a 
complex branch with $\theta(\gamma)=0$ emanating from the point $\gamma = -2\pi T$ and saturating around ${\rm Re}\gamma \sim -\pi T$. As $\Omega_0/T$ increases, this saturation drifts towards ${\rm Re \gamma}=-2\pi T$, until at some point it loses relevance due to it being surpassed by the bottleneck $\gamma_*$.

Fig.~\ref{figure 2}(bottom) shows the corresponding solutions $\gamma(g^2/T)$ extracted from the real and complex branches, for the example with $\Omega_0/T=1$. In this case, the complex branch ``beats'' the real branch above $g^2/T^2\gtrsim4$. Figure~\ref{figure 3} of the main text was created by compiling such plots for many values of $\Omega_0/T$. As can be seen from the behaviour in Fig.~\ref{figure 2}(bottom), the diffusive branch, which is linear in $g^2/T^2$, gives way to either the complex branch or the $\gamma_*$ bottleneck at larger $g/T$. Which one of these dominates is determined by the one which saturates at a smaller (in magnitude) real part, with the bottleneck taking over slightly above $\Omega_0/T=1$.

\end{document}